\documentclass[a4paper,fleqn,usenatbib]{mnras}

\usepackage[T1]{fontenc}
\usepackage{ae,aecompl}
\usepackage[dvipdfmx]{graphicx}
\usepackage{amsmath,amssymb}
\usepackage{textgreek}
\usepackage{booktabs}
\newcommand{\abs}[1]{\lvert #1\rvert}

\newcommand{\ur}[1]{\,\mathrm{#1}}
\usepackage{siunitx}
\usepackage{mhchem}

\defcitealias{Taniguchi2018}{Paper~I}
\defcitealias{Jian2020}{Paper~II}

\title[$T_{\mathrm{eff}}$ of RSGs using LDRs of \ion{Fe}{i} lines in \textit{YJ} bands]{Effective temperatures of red supergiants estimated from line-depth ratios of iron lines in the \textit{YJ} bands, $0.97\text{--}1.32\,\text{\textmu m}$}

\author[D. Taniguchi et al.]{Daisuke Taniguchi,$^{1}$\thanks{E-mail: taniguchi@astron.s.u-tokyo.ac.jp} Noriyuki Matsunaga,$^{1,2}$\thanks{E-mail: matsunaga@astron.s.u-tokyo.ac.jp} Mingjie Jian,$^{1}$ Naoto Kobayashi,$^{3,4,2}$ 
\newauthor Kei Fukue,$^{2}$ Satoshi Hamano,$^{2,5}$ Yuji Ikeda,$^{2,6}$ Hideyo Kawakita,$^{2,7}$ Sohei Kondo,$^{2,4}$
\newauthor Shogo Otsubo,$^{2}$ Hiroaki Sameshima,$^{3}$ Keiichi Takenaka$^{8}$ and Chikako Yasui$^{5}$ \\
$^{1}$Department of Astronomy, The University of Tokyo, 7-3-1 Hongo, Bunkyo-ku, Tokyo 113-0033, Japan \\
$^{2}$Laboratory of Infrared High-resolution spectroscopy (LiH), Koyama Astronomical Observatory, Kyoto Sangyo University, Motoyama, \\ Kamigamo, Kita-ku, Kyoto 603-8555, Japan \\
$^{3}$Institute of Astronomy, The University of Tokyo, 2-21-1 Osawa, Mitaka, Tokyo 181-0015, Japan \\
$^{4}$Kiso Observatory, The University of Tokyo, 10762-30 Mitake, Kiso-machi, Kiso-gun,Nagano 397-0101, Japan \\
$^{5}$National Astronomical Observatory of Japan, 2-21-1 Osawa, Mitaka, Tokyo 181-8588, Japan \\
$^{6}$Photocoding, 460-102 Iwakura-Nakamachi, Sakyo-ku, Kyoto 606-0025, Japan \\
$^{7}$Department of Astrophysics and Atmospheric Sciences, Faculty of Sciences, Kyoto Sangyo University, Motoyama, Kamigamo, \\ Kita-ku, Kyoto 603-8555, Japan \\
$^{8}$Division of Science, Graduate School, Kyoto Sangyo University, Motoyama, Kamigamo, Kita-ku, Kyoto 603-8555, Japan}

\date{Accepted XXX. Received YYY; in original form ZZZ}
\pubyear{2020}

\begin{document}
\label{firstpage}
\pagerange{\pageref{firstpage}--\pageref{lastpage}}
\maketitle

\begin{abstract}
Determining the effective temperatures~($T_{\mathrm{eff}}$) of red supergiants~(RSGs) observationally is important in many fields of stellar physics and galactic astronomy, yet some significant difficulties remain due to model uncertainty originating majorly in the extended atmosphere of RSGs. 
Here we propose the line-depth ratio~(LDR) method in which we use only \ion{Fe}{i} lines. 
As opposed to the conventional LDR method with lines of multiple species involved, the LDR of this kind is insensitive to the surface gravity effects and expected to circumvent the uncertainty originating in the upper atmosphere of RSGs. 
Therefore, the LDR--$T_{\mathrm{eff}}$ relations that we calibrated empirically with red giants may be directly applied to RSGs, though various differences, e.g., caused by the three-dimensional non-LTE effects, between the two groups of objects need to be kept in mind. 
Using the near-infrared \textit{YJ}-band spectra of nine well-known solar-metal red giants observed with the WINERED high-resolution spectrograph, we selected $12$ pairs of \ion{Fe}{i} lines least contaminated with other lines. 
Applying their LDR--$T_{\mathrm{eff}}$ relations to ten nearby RSGs, the resultant $T_{\mathrm{eff}}$ with the internal precision of $30\text{--}70\ur{K}$ shows good agreement with previous observational results assuming one-dimensional LTE and with Geneva's stellar evolution model.
We found no evidence of significant systematic bias caused by various differences, including those in the size of the non-LTE effects, between red giants and RSGs except for one line pair which we rejected because the non-LTE effects may be as large as $\sim 250\ur{K}$. 
Nevertheless, it is difficult to evaluate the systematic bias, and further study is required, e.g., with including the three-dimensional non-LTE calculations of all the lines involved. 
\end{abstract}


\section{Introduction}

\subsection{Effective temperatures of red supergiants}\label{sec:IntroRSGTeff}

The effective temperature~($T_{\mathrm{eff}}$) of the red supergiant~(RSG) is one of the essential parameters in many fields of stellar physics, e.g., the theory of stellar evolution~\citep[e.g.][]{Massey2003b,Ekstrom2012,Choi2016} and theory of the maximum initial mass of the type-II supernova progenitor~\citep[e.g.][]{Fraser2011,Smartt2015}.
Moreover, the accuracy of $T_{\mathrm{eff}}$ directly affects that of chemical abundances of RSGs, which could trace the abundance distribution of young stars in the Milky Way and nearby galaxies, leading to the galactic evolution theory~\citep{Patrick2017,AlonsoSantiago2019,Origlia2019}. 
Hence importance of accurately determining $T_{\mathrm{eff}}$ of RSGs with observations cannot be overstated.

However, the accuracy of the reported $T_{\mathrm{eff}}$ of RSGs is still under debate. 
Interferometry and lunar occultation are the two simplest ways to measure $T_{\mathrm{eff}}$ of nearby stars, but these methods are subject to uncertainties in some or all of the following three issues: (1)~limb darkening~\citep{Dyck2002,Chiavassa2009}, (2)~the complex extended envelope called MOLsphere~\citep{Tsuji2006,Montarges2014} and (3)~interstellar reddening~\citep{Massey2005,Walmswell2012}. 
A more recent and improved way is to measure the stellar radius at continuum wavelengths to circumvent the effects of the MOLsphere with the spectro-interferometric technique~\citep[e.g.][]{Ohnaka2013,ArroyoTorres2013}; however, the result is still affected by limb darkening and possibly uncertainty in interstellar reddening unless it is negligible. 

Another type of method to estimate $T_{\mathrm{eff}}$ of the RSG is to model-fit molecular bands in relatively low-resolution spectra in the optical~\citep[e.g.][]{Scargle1979,Oestreicher1998}. 
The reliability of this approach has been improved thanks to the advancements of the model of the stellar atmosphere such as, most notably, the MARCS model with improved handling of molecular blanketing~\citep{Gustafsson2008}. 
The most comprehensive work of this approach was presented by \citet{Levesque2005} for RSGs in the Milky Way, in which they made use of the optical \ce{TiO} bands~\citep[see also][where RSGs in other galaxies were studied with the same method]{Levesque2006,Massey2009,Massey2016}; their method is hereafter referred to as the \ce{TiO} method. 

A similar approach is to compare photometric colours of RSGs with the prediction of the MARCS model~\citep[e.g.][]{Levesque2005,Drout2012,Neugent2012}, in which molecular lines are taken into account to calculate the spectral energy distribution~(SED) of the RSG. 
This approach tends to give a lower $T_{\mathrm{eff}}$ in literature than the \ce{TiO} method by up to $\sim 200\ur{K}$~\citep{Levesque2005,Levesque2006}, and hence there remain systematic uncertainties in fitting molecular absorption in optical spectra~\citep{Davies2013}. 

Alternatively some authors estimated $T_{\mathrm{eff}}$ and spectral types of RSGs from molecular \ce{CO}, \ce{H2O}, \ce{OH} and/or \ce{CN} lines in near-infrared spectra~\citep[e.g.][]{Carr2000,Cunha2007,Origlia2019}. 
However, \citet{Lancon2007} and \citet{Lancon2010} found that the strengths of the \ce{CN} molecular bands and the ratios of the strengths of the \ce{CO} bandheads could not be reproduced with modern models of the static photosphere for the CNO abundances as predicted by stellar evolution models and a microturbulent velocity~($v_{\mathrm{micro}}$) of $2\ur{\si{km.s^{-1}}}$. 

The temperature estimated on the basis of molecular bands are affected more or less by a few factors, including chemical abundances and discrepancy between the atmosphere of the real star and that estimated with a simplified static model assuming one-dimensional local thermodynamic equilibrium~(LTE) without MOLsphere. 
 For example, \citet{Chiavassa2011b} and \citet{Davies2013} found that the \ce{TiO}-band strength depends not only on $T_{\mathrm{eff}}$ and the luminosity but also on the temperature structure, which is significantly affected by granulation. 

A promising way to circumvent the problems raised so far is to use signals like weak atomic lines from the photosphere defined by the optical continuum or its outside close vicinity, given that they are expected to be less affected by some of the above-mentioned factors~\citep{Davies2013,Tabernero2018}. 
A few works have presented $T_{\mathrm{eff}}$ of RSGs estimated on the basis of photospheric signals, e.g., the SED of wavelengths less contaminated by molecular lines~\citep[the SED method;][]{Davies2013}, equivalent widths of some strong atomic lines in the \textit{J} band~\citep[the \textit{J}-band technique;][]{Davies2015} and some atomic lines around the Calcium triplet~\citep{Tabernero2018}. 
In these three works $T_{\mathrm{eff}}$ of common RSGs in the Magellanic Clouds were derived. 
Nevertheless some systematic offsets are apparent between their results, possibly due to chemical abundances, contamination from some molecular lines to the signals and/or some effect of strong lines which is in part formed in upper layers of the atmosphere. 
Moreover, the departure from the 1D LTE condition may be important in RSGs~\citep[e.g.][]{Bergemann2012b,Kravchenko2019}. 
Most of the above works assumed the 1D LTE, and this shortcoming could contribute to the offsets. 

Considering these discrepancies and the aforementioned factors that possibly add some systematic error to the temperatures estimated in most of the past methods, we here take a different method that satisfies the following three conditions. 
First, the method should use only relatively weak atomic lines that are not severely contaminated by molecular lines. 
Such lines do not originate in atmospheric layers far above the photosphere, where the temperature structure is not well constrained and moreover is expected to show variability. 
Second, the method should be independent of stellar parameters, abundances and reddening of both interstellar and circumstellar origins as much as possible. 
Third, ideally, the method should be independent of uncertainties in theoretical models, in particular with regard to the factors originating from the parameters in the upper atmosphere layers. 
Therefore, we use the `line-depth ratio'~(LDR) method, which relies on the empirical calibration of the relations between the LDR and $T_{\mathrm{eff}}$ and satisfies the above-mentioned three requirements. In the next subsection, we explain what the LDR and LDR-method are and review relevant past studies.

\subsection{Line-depth ratio as an indicator of the effective temperature}\label{sec:IntroLDR}

In cool stars~(with a temperature of the solar temperature or lower), the depths of low-excitation lines of neutral atoms tend to be sensitive to $T_{\mathrm{eff}}$ whereas those of high-excitation lines are relatively insensitive~\citep{Gray2008a}. Therefore, the ratios of the depths between the low- and high-excitation lines (that is, the LDRs) are good temperature indicators~\citep[and references therein]{Gray1991,Teixeira2016,LopezValdivia2019}. 
The basic procedure is as follows: (1)~search high-resolution spectra for the line pairs whose depth ratios are sensitive to $T_{\mathrm{eff}}$, (2)~establish the empirical relations between LDRs and $T_{\mathrm{eff}}$ for stars with well-determined $T_{\mathrm{eff}}$ and (3)~apply the relations to target objects to determine their $T_{\mathrm{eff}}$. 

We focus on the near-infrared \textit{Y} and \textit{J} bands. 
These bands host a relatively small number of molecular lines in the RSG spectra~\citep{Coelho2005,Davies2010}, and we can find sufficiently many atomic lines that are not contaminated by molecular lines~(see \autoref{ssec:LineSelection} for detail and real examples). 
 \citet[hereafter \citetalias{Taniguchi2018}]{Taniguchi2018} recently investigated LDRs in the \textit{Y} and \textit{J} bands of ten red giants with well-determined effective temperatures~($3700<T_{\mathrm{eff}}<5400\ur{K}$). 
They calibrated $81$ LDR--$T_{\mathrm{eff}}$ relations with the neutral atomic lines of various elements, and found that a precision of $\pm 10\ur{K}$ is achievable in the best cases, i.e., high-resolution spectra of early M-type red giants with a good signal-to-noise ratio~(S/N). 
This precision rivals those achieved in the previous results in which the LDRs in optical high-resolution spectra were employed~\citep[e.g.][]{Kovtyukh2006}. 

Although \citetalias{Taniguchi2018} gave well-defined LDR--$T_{\mathrm{eff}}$ relations for solar-metal red giants, some significant systematic uncertainty remains in their work, where a simple relation between the LDR and $T_{\mathrm{eff}}$ was assumed to hold universally. 
In reality, LDRs depend on other stellar parameters than $T_{\mathrm{eff}}$, even though $T_{\mathrm{eff}}$ is the dominant parameter that determines the LDR of a neutral atom. 
\citet{Jian2019}, for example, demonstrated that the LDR--$T_{\mathrm{eff}}$ relations in the \textit{H} band depend on the metallicity~([Fe/H]) by $100\text{--}800\ur{K/dex}$. 
Also, \citetalias{Taniguchi2018} suggested some effect of [Fe/H] and abundance ratios on their LDR--$T_{\mathrm{eff}}$ relations. 
In this work, we limit the sample to objects with [Fe/H] around solar~(\autoref{ssec:sample}) to circumvent the problem of the [Fe/H] dependence. 

Surface gravity effect on LDRs has been also detected. 
\citet[hereafter \citetalias{Jian2020}]{Jian2020} compared the LDRs of the line pairs, reported by \citetalias{Taniguchi2018}, between $20$ dwarfs, $25$ giants and $18$ supergiants and reported the LDRs' dependency on the surface gravity. 
This dependency is explained with difference in the ionization stages among the elements involved in the line pairs. 
All the past attempts to derive a number of relations utilized any combination of usable species among selected elements; accordingly, their estimates might be significantly affected by the surface gravity. 
The LDRs of two lines with common species are, in contrast, insensitive to the surface gravity. 
This kind of LDRs is also insensitive to the chemical abundance ratios. 
Therefore, the LDR--$T_{\mathrm{eff}}$ relations derived on the basis of line pairs of the same species which are calibrated with the spectra of red giants are, when applied to RSGs, expected to yield $T_{\mathrm{eff}}$ without introducing large systematic errors. 
Specifically, we use \ion{Fe}{i} lines in this work because they are the only species that gives a sufficient number of the lines, hence a sufficient number of the line pairs, with which well-constrained LDR--$T_{\mathrm{eff}}$ relations can be derived. 
Nevertheless, the difference in the size of the 3D non-LTE effect between red giants and RSGs could introduce systematic errors to $T_{\mathrm{eff}}$ of RSGs obtained with our method, and should be examined. 

In this work, using the high-resolution spectra in the near-infrared \textit{Y} and \textit{J} bands of solar-metal red giants with well-determined stellar parameters observed with the WINERED spectrograph~(\autoref{sec:Obs}), we construct a set of many empirical LDR--$T_{\mathrm{eff}}$ relations with only \ion{Fe}{i} lines for the first time~(\autoref{sec:ConstructingLDR}). 
Then, we apply these relations to the WINERED spectra of nearby RSGs and determine their $T_{\mathrm{eff}}$ in an unprecedented accuracy as non-spectro-interferometric measurements~(\autoref{sec:TeffRSGs}).

\section{Observations and Reduction}\label{sec:Obs}

\subsection{Sample}\label{ssec:sample}

We consider two groups of targets for this work. 
The first group consists of nine well-known red giants and are used for calibration of the LDR--$T_{\mathrm{eff}}$ relations in this work. 
Five of these stars (\textepsilon ~Leo, Pollux, \textmu ~Leo, Aldebaran and \textalpha ~Cet) were selected from \textit{Gaia} benchmark stars~\citep{BlancoCuaresma2014}, whose $T_{\mathrm{eff}}$ were well constrained with interferometry by \citet{Heiter2015}. 
The other four stars are well-studied nearby red giants selected from the MILES sample~\citep{SanchezBlazquez2006}. 
The temperatures of these stars were determined using the ULySS program~\citep{Koleva2009} by \citet{Prugniel2011} with the MILES empirical spectral library as a reference. 
The temperature scale of \citet{Prugniel2011} relies on the compilation of literature stellar parameters, for which most spectroscopic works assumed the 1D LTE condition, and we have checked the consistency between the temperature scale of \citet{Prugniel2011} and that of \citet{Heiter2015} as follows. 
There are $13$ stars\footnote{The $13$ stars common in \citet{Prugniel2011} and \citet{Heiter2015} have HD numbers: 6582, 10700, 22049, 22879, 49933, 102870, 201091~(dwarfs), 18907, 23249, 121370~(subgiants), 29139, 85503 and 220009~(giants). } with [Fe/H] higher than $-1.0\ur{dex}$ that were investigated in both papers. 
The temperatures from the two studies differ by $28\ur{K}$ on average with the unbiased standard deviation of the differences being $68\ur{K}$, which is comparable with the combined measurement errors. 
Thus, the temperature scale of \citet{Prugniel2011} is the same as that of \citet{Heiter2015} within the errors, and we make use of the temperatures by \citet{Prugniel2011}. 
The stellar parameters of these nine red giants are summarized in \autoref{table:BenchmarkAtmos}. 
We also adopted elemental abundances from \citet{Jofre2015} if available. 
We note that these red giants were also used in \citetalias{Taniguchi2018} to establish the LDR--$T_{\mathrm{eff}}$ relations from pairs of lines with all possible combinations of the selected elements. 
The parameters $T_{\mathrm{eff}}$ of the red giants range from $3700\ur{K}$ to $5400\ur{K}$ and are used in the calibration of the LDR--$T_{\mathrm{eff}}$ relations~(\autoref{ssec:LinePairSelection}). 
The other stellar parameters are used only to check blending of absorption lines and to select the useful lines~(\autoref{ssec:LineSelection}). 

The second group of the targets consists of ten nearby RSGs and is the main target group in this work. 
Their [Fe/H] are close to solar~\citep{Luck1989,McWilliam1990,Wu2011}, as expected for young stars in the solar neighbourhood. 
The effective temperatures $T_{\mathrm{eff}}$ of all these ten RSGs were estimated by \citet{Levesque2005} using the \ce{TiO} method~($3600\ur{K}<T_{\mathrm{eff}}<4000\ur{K}$); thus we can compare our derived $T_{\mathrm{eff}}$ values and theirs. 
Four of our sample, NO Aur, V809 Cas, 41 Gem and \textxi \ Cyg, were selected from the sample in \citet{Kovtyukh2007} and they were used also by \citetalias{Jian2020} except V809 Cas to investigate the LDR--$T_{\mathrm{eff}}$ relations of supergiants~($3900<T_{\mathrm{eff}}<6300\ur{K}$), where lines of various elements were employed. 

\begin{table*}
\caption{The stellar parameters of the calibrating nine red giants. Full-width at half maximum~(FWHM) indicates the broadening width assuming Gaussian function that includes instrumental~($\sim 10.7\ur{\si{km.s^{-1}}}$), macroturbulent and rotational broadenings. }
\label{table:BenchmarkAtmos}
\begin{tabular}{lccclc}\toprule 
Name & $T_{\mathrm{eff}}\ \mathrm{[K]}$ & [Fe/H] [dex] & $\log g\ \mathrm{[dex]}$ & $v_{\mathrm{micro}}$ [\si{km.s^{-1}}] & FWHM [\si{km.s^{-1}}]${}^{\text{[7]}}$ \\ \midrule 
\textepsilon \ Leo & $5398\pm 31$${}^{\text{[2]}}$ & $-0.06\pm 0.04$${}^{\text{[2]}}$ & $2.02\pm 0.08$${}^{\text{[2]}}$ & $1.61$${}^{\text{[6]}}$ & $15.91$ \\
\textkappa \ Gem & $5029\pm 47$${}^{\text{[2]}}$ & $-0.01\pm 0.05$${}^{\text{[2]}}$ & $2.61\pm 0.12$${}^{\text{[2]}}$ & $1.47$${}^{\text{[6]}}$ & $12.49$ \\
\textepsilon \ Vir & $4983\pm 61$${}^{\text{[1]}}$ & $+0.10\pm 0.16_{}$${}^{\text{[4]}}$ & $2.77\pm 0.02$${}^{\text{[1]}}$ & $1.39\pm 0.25$${}^{\text{[5]}}$ & $12.42$ \\
Pollux & $4858\pm 60$${}^{\text{[1]}}$ & $+0.08\pm 0.16$${}^{\text{[4]}}$ & $2.90\pm 0.08$${}^{\text{[1]}}$ & $1.28\pm 0.21$${}^{\text{[5]}}$ & $11.74$ \\
\textmugreek \ Leo & $4470\pm 40$${}^{\text{[3]}}$ & $+0.20\pm 0.15$${}^{\text{[4]}}$ & $2.51\pm 0.11$${}^{\text{[1]}}$ & $1.28\pm 0.26$${}^{\text{[5]}}$ & $12.56$ \\
Alphard & $4171\pm 52$${}^{\text{[2]}}$ & $+0.08\pm 0.07$${}^{\text{[2]}}$ & $1.56\pm0.20$${}^{\text{[2]}}$ & $1.66$${}^{\text{[6]}}$ & $12.71$ \\
Aldebaran & $3882\pm 19$${}^{\text{[3]}}$ & $-0.42\pm 0.17$${}^{\text{[4]}}$ & $1.11\pm 0.19$${}^{\text{[1]}}$ & $1.63\pm 0.30$${}^{\text{[5]}}$ & $13.19$ \\
\textalpha \ Cet & $3796\pm 65$${}^{\text{[1]}}$ & $-0.50\pm 0.47$${}^{\text{[4]}}$ & $0.68\pm 0.23$${}^{\text{[1]}}$ & $1.77\pm 0.40$${}^{\text{[5]}}$ & $12.87$ \\
\textdelta \ Oph & $3783\pm 20$${}^{\text{[2]}}$ & $-0.03\pm 0.06$${}^{\text{[2]}}$ & $1.45\pm 0.19$${}^{\text{[2]}}$ & $1.66$${}^{\text{[6]}}$ & $12.67$ \\ \bottomrule 
\end{tabular}

References: 
[1]~\citet{Heiter2015}; 
[2]~\citet{Prugniel2011}; 
[3]~Weignted mean of values in \citet{Heiter2015} and \citet{Prugniel2011}; 
[4]~\citet{Heiter2015} converted to Solar metallicity of $A(\ce{Fe})_{\odot }=7.50\ur{dex}$~\citep{Asplund2009}; 
[5]~\citet{Jofre2015}; 
[6]~Estimated using the relation between $\log g$ and $v_{\mathrm{micro}}$ for ASPCAP DR13~\citep{Holtzman2018}; 
[7]~Measured by the fitting of several isolated lines. 
\end{table*}

\begin{table*}
\centering 
\caption{Observation log of our sample stars. Spectral types are taken from SIMBAD~\citep{Wenger2000} on 2020 April 26. }
\label{table:ObsLog}
\begin{tabular}{lrl lrl l}\toprule 
\multicolumn{3}{c}{Object} & \multicolumn{3}{c}{Telluric Standard} & Obs. Date \\ \cmidrule(lr){1-3}\cmidrule(lr){4-6}
Name & HD & Sp. Type & Name & HD & Sp. Type & \\ \midrule 
\multicolumn{7}{l}{Nine well-known red giants} \\
\textepsilon \ Leo & 84441 & G1IIIa & 21 Lyn & 58142 & A0.5Vs & 2014 Jan 23 \\
\textkappa \ Gem & 62345 & G8III--IIIb & HR 1483 & 29573 & A0V & 2013 Dec 8 \\
\textepsilon \ Vir & 113226 & G8III--IIIb & b Vir & 104181 & A0V & 2014 Jan 23 \\
Pollux & 62509 & K0IIIb & HIP 58001 & 103287 & A0Ve+K2V & 2013 Feb 28 \\
\textmugreek \ Leo & 85503 & K2IIIbCN1Ca1 & HIP 76267 & 139006 & A1IV & 2013 Feb 23 \\
Alphard & 81797 & K3IIIa & HR 1041 & 21402 & A2V & 2013 Nov 30 \\
Aldebaran & 29139 & K5+III & HIP 28360 & 40183 & A1IV--Vp & 2013 Feb 24 \\
\textalpha \ Cet & 18884 & M1.5IIIa & omi Aur & 38104 & A2VpCr & 2013 Nov 30 \\
\textdelta \ Oph & 146051 & M0.5III & b Vir & 104181 & A0V & 2014 Jan 23 \\ \midrule 
\multicolumn{7}{l}{Ten nearby RSGs} \\
\textzeta \ Cep & 210745 & K1.5Ib & HR 6432 & 156653 & A1V & 2015 Aug 8 \\
41 Gem & 52005 & K3--Ib & HR 922 & 19065 & B9V & 2015 Oct 28 \\
\textxi \ Cyg & 200905 & K4.5Ib--II & 39 UMa & 92728 & A0III & 2016 May 14 \\
V809 Cas & 219978 & K4.5Ib & c And & 14212 & A0V & 2015 Oct 31 \\
V424 Lac & 216946 & K5Ib & HR 8962 & 222109 & B8V & 2015 Jul 30 \\
\textpsi ${}^{1}$ Aur & 44537 & K5--M1Iab--Ib & HIP 53910 & 95418 & A1IVps & 2013 Feb 22 \\
TV Gem & 42475 & M0--M1.5Iab & 50 Cnc & 74873 & A1Vp & 2016 Jan 19 \\
BU Gem & 42543 & M1--M2Ia--Iab & 50 Cnc & 74873 & A1Vp & 2016 Jan 19 \\
Betelgeuse & 39801 & M1--M2Ia--Iab & HIP 27830 & 39357 & A0V & 2013 Feb 22 \\
NO Aur & 37536 & M2Iab & HR 922 & 19065 & B9V & 2015 Oct 28 \\
\bottomrule 
\end{tabular}
\end{table*}

\subsection{Observations}\label{ssec:InstAndObs}

All the objects were observed using the near-infrared high-resolution spectrograph WINERED installed on the Nasmyth platform of the $1.3\ur{m}$ Araki Telescope at Koyama Astronomical Observatory of Kyoto Sangyo University in Japan~\citep{Ikeda2016}. 
We used the WINERED WIDE mode to collect spectra covering a wavelength range from $0.90$ to $1.35\,\text{\textmugreek m}$~(\textit{z$^{\prime }$}, \textit{Y} and \textit{J} bands) with a spectral resolution of $R\sim 28,000$. 
We selected the nodding pattern of A--B--B--A or O--S--O. 
All our targets are bright~($-3.0\leq J\leq 3.0\ur{mag}$), and the total integration time for each target within the slit ranged between $3\text{--}240\ur{sec}$, with which a S/N per pixel of $100$ or higher was achieved. 
Telluric standard stars~\citep[slow-rotating A0V stars in most cases; see][]{Sameshima2018} were also observed, the spectra of which were used to subtract the telluric absorption. 
\autoref{table:ObsLog} summarizes the observation log. 

As in \citetalias{Taniguchi2018}, we utilized the echelle orders 57th--52nd~(\textit{Y} band; $0.97\text{--}1.09\,\text{\textmu m}$) and 48th--43rd~(\textit{J} band; $1.15\text{--}1.32\,\text{\textmu m}$) only among the orders 61st--42nd because stellar atomic lines in the unselected orders are severely contaminated by other lines.
The orders 61st--59th, 50th--49th and 42nd are dominated by telluric lines in our spectra collected in Kyoto~\citep[see Figure 4 in][]{Sameshima2018}. 
The orders 58th and 51st are dominated by stellar \ce{CN} molecular lines of the (1--0) bandhead and those of the (0--0) band of the \ce{CN} $A{}^{2}\Pi $--$X{}^{2}\Sigma ^{+}$ red system, respectively~\citep[e.g.][]{Kurucz2011,Brooke2014}. 
These two orders are heavily contaminated also by telluric lines in the spectra taken in summer. 
The wavelength ranges of the individual orders of WINERED are given in \autoref{table:SNs}. 

\begin{figure*}
\centering 
\includegraphics[width=\textwidth ]{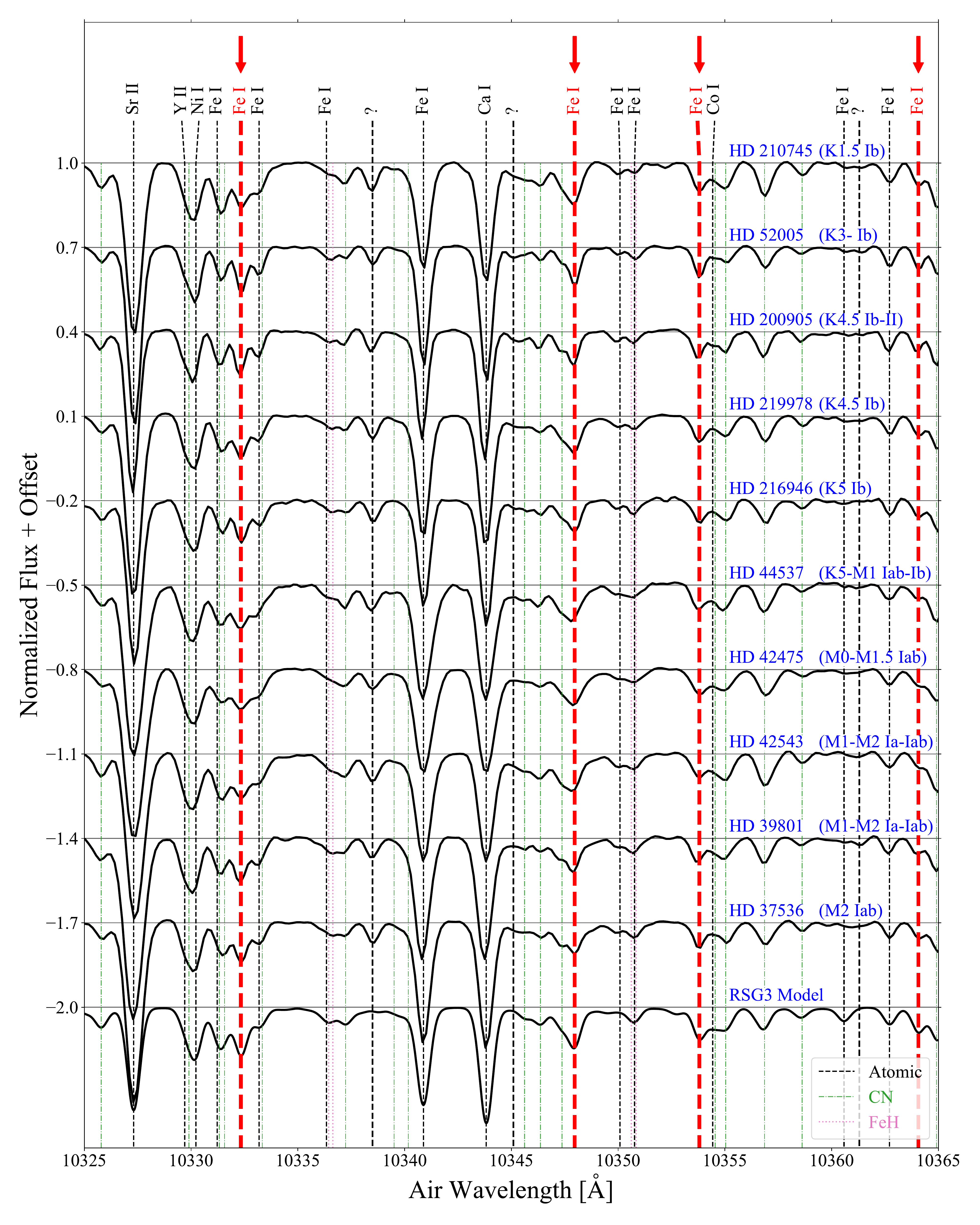}
\caption{A part of the reduced WINERED spectra and the model RSG3 spectrum in the echelle order 54th with some identified lines labelled. Four \ion{Fe}{i} lines indicated by red arrows are used for the final set of the LDR line pairs~(\autoref{table:LDRFePairs}). We have not identified the origins of the three stellar lines labelled with `?', among which the line at $10338.5\,\text{\AA }$ is catalogued in the list of unidentified lines by \citet{Matsunaga2020}. }
\label{fig:BetelgeuseSpec}
\end{figure*}

\subsection{Data reduction}\label{sec:Reduction}

The initial steps of the spectral reduction were performed with the WINERED data-reduction pipeline~(Hamano et al. in preparation). 
Some details of the pipeline are given in Section 3.1 in~\citetalias{Taniguchi2018}. 
Then, the telluric absorption lines were removed using the observed spectra of A0V stars after their intrinsic lines had been removed with the method described in \citet{Sameshima2018}, except for the 55th--53rd orders~($1.01$ to $1.07\,\text{\textmugreek m}$) of the objects taken in winter, in which almost no significant telluric lines were present. 
The radial velocities were measured and corrected by comparing the observed and synthesized spectra. 
Finally the continuum was re-normalized and the realistic S/N was estimated as we describe in the following subsections. 
An example of the reduced spectra of RSGs is presented in \autoref{fig:BetelgeuseSpec}.

\subsubsection{Continuum Normalization}\label{sssec:continuum}

The nominal continuum normalization was automatically performed in the data-reduction pipeline, but the automatic normalization often yields unsatisfactory results, especially for line-rich stars like RSGs. 
Therefore, we further optimized the normalizations for our spectra in the following procedure.
First, we prepared the target spectra with the telluric absorption lines subtracted and with the wavelength-scale adjusted to the one in the air at rest. 
Second, we selected 	a set of continuum regions for each group of the red giants and RSGs, where the spectra are not significantly affected by the stellar lines. 
For the nine red giants, the continuum regions were selected with visual inspection. 
In contrast, we searched for the continuum regions of the RSGs by choosing the wavelength ranges with the depths from unity smaller than $0.3\%$ in the model spectrum with the stellar parameters of `RSG3' in \autoref{table:ImaginaryRSGAtmos}~(see \autoref{sec:ConstructingLDR} about the spectral synthesis). 
These steps left a few tens of evenly distributing continuum segments in each order. 
Finally, we normalized the continuum of each order of each target star with the interactive mode of the \textsc{IRAF} \texttt{continuum} task, in which we mainly used the selected continuum regions but sometimes further combining some continuum regions visually selected for each order of each star. 

\begin{table}
\centering 
\caption{Stellar parameters used to simulate spectra of red giants and RSGs; we use these spectra for various purposes~(see text). We assumed the solar chemical abundance pattern~\citep{Asplund2009}. }
\label{table:ImaginaryRSGAtmos}
\begin{tabular}{lllll}\toprule 
Model & RSG1 & RSG2 & RSG3 & RGB1 \\ \midrule 
Sp. Type & K3I & M1.5I & M0I & M0.5III \\
$T_{\mathrm{eff}}$ [K] & $4000$ & $3700$ & $3850$ & $3850$ \\
{[Fe/H]} [dex] & $0.0$ & $0.0$ & $0.0$ & $0.0$ \\
$\log g$ [dex] & $0.5$ & $0.0$ & $0.25$ & $1.0$ \\
$v_{\mathrm{micro}}$ [\si{km.s^{-1}}] & $1.8$ & $2.5$ & $2.15$ & $1.7$ \\
FWHM [\si{km.s^{-1}}] & $16$ & $21$ & $18.5$ & $13$ \\ \bottomrule 
\end{tabular}
\end{table}

\subsubsection{Signal-to-Noise Ratios of Telluric-Subtracted Spectra}\label{sssec:SNEstimateReal}

We need to know the errors in line depth for calibrating the LDR--$T_{\mathrm{eff}}$ relations and for estimating $T_{\mathrm{eff}}$. 
The depth errors are considered to originate in three sources: (1)~statistical noises in the telluric-subtracted spectra, which can be estimated with a combination of S/N of the target and telluric standard spectra~($200$ per pixel or higher in most of the cases), (2)~residuals in the telluric subtraction and (3)~systematic errors caused by slightly wrong continuum placement. 
Though the errors introduced by the first source can be estimated using a certain method, e.g., the one by \citet[see their Section 3.2]{Fukue2015}, those by the subsequent sources cannot be estimated in a straightforward manner. 

\begin{table*}
\centering 
\caption{`Effective' S/N per pixel of the reduced spectra of RSGs in each echelle order~(57th to 52nd in the \textit{Y} band and 48th to 43rd in \textit{J}) and the wavelength coverage, $\lambda _{\mathrm{min}}<\lambda _{\mathrm{air}}<\lambda _{\mathrm{max}}$. }
\label{table:SNs}
\begin{tabular}{lrrrrrrrrrrrr}\toprule 
Object & m57 & m56 & m55 & m54 & m53 & m52 & m48 & m47 & m46 & m45 & m44 & m43 \\ \midrule 
\textzeta \ Cep & $75$ & $112$ & $70$ & $86$ & $108$ & $113$ & $57$ & $56$ & $68$ & $120$ & $80$ & $73$ \\
41 Gem & $55$ & $149$ & $123$ & $119$ & $107$ & $117$ & $50$ & $61$ & $74$ & $147$ & $63$ & $81$ \\
\textxi \ Cyg & $98$ & $151$ & $92$ & $148$ & $138$ & $84$ & $60$ & $55$ & $106$ & $138$ & $93$ & $61$ \\
V809 Cas & $103$ & $122$ & $58$ & $100$ & $127$ & $91$ & $60$ & $64$ & $95$ & $128$ & $99$ & $65$ \\
V424 Lac & $57$ & $59$ & $68$ & $99$ & $80$ & $69$ & $54$ & $57$ & $83$ & $147$ & $58$ & $66$ \\
\textpsi ${}^{1}$ Aur & $109$ & $138$ & $105$ & $144$ & $71$ & $64$ & $86$ & $135$ & $75$ & $167$ & $81$ & $103$ \\
TV Gem & $89$ & $98$ & $107$ & $105$ & $82$ & $68$ & $66$ & $58$ & $78$ & $191$ & $51$ & $57$ \\
BU Gem & $73$ & $69$ & $77$ & $75$ & $81$ & $73$ & $85$ & $53$ & $57$ & $109$ & $70$ & $54$ \\
Betelgeuse & $49$ & $53$ & $55$ & $87$ & $117$ & $75$ & $89$ & $89$ & $60$ & $146$ & $64$ & $60$ \\
NO Aur & $66$ & $150$ & $59$ & $85$ & $133$ & $89$ & $46$ & $54$ & $70$ & $123$ & $85$ & $72$ \\
\midrule 
$\lambda _{\mathrm{min}}$ [\textmu m] & $0.976$ & $0.993$ & $1.011$ & $1.029$ & $1.049$ & $1.069$ & $1.156$ & $1.181$ & $1.206$ & $1.233$ & $1.260$ & $1.289$ \\
$\lambda _{\mathrm{max}}$ [\textmu m] & $0.993$ & $1.011$ & $1.029$ & $1.049$ & $1.069$ & $1.089$ & $1.181$ & $1.206$ & $1.233$ & $1.260$ & $1.289$ & $1.319$ \\
\bottomrule 
\end{tabular}
\end{table*}

In order to estimate realistic errors in line depths including all the above-mentioned three sources, we considered the `effective' S/N per pixel of the telluric-subtracted spectra as follows. 
For this, we used the ready-to-use target spectra with normalized continuum. 
First, we measured the deviations, $\delta _{i}$, from unity of each pixel, $i$, within the continuum regions selected in \autoref{sssec:continuum}. 
The pixels $\{i\}$ that satisfy $\abs{\delta _{i}}>0.05$ might not be in real continuum regions, and thus were excluded in the subsequent analysis. 
Of course, such a threshold cannot be used for low-S/N spectra, but our spectra are sufficiently high `effective' S/N, $\sim 50$ at least. 
We note that some of such pixels might be due to the absorption lines in the observed RSG spectra that were not predicted in the synthesized spectra, and others might be induced by the bad continuum normalization. 
The `effective' S/N were then estimated according to, after three iterations of three-sigma clipping,
\begin{equation}
\text{S/N}=\left[\frac{\sum _{i}{\delta _{i}}^{2}}{N_{\mathrm{pixel}}-1}\right]^{-1/2}\text{,}
\end{equation}
where $N_{\mathrm{pixel}}$ is the number of the used pixels. 
We note that the sigma-clipping may underestimate the error, but by only $2\%$ at most, providing that $\delta _{i}$ follows the Gaussian distribution. 
The resultant S/N of the RSGs are summarized in \autoref{table:SNs}, whereas those for the red giants are found in Table 2 in \citetalias{Taniguchi2018}. 
The reciprocal of the S/N measured here are henceforth regarded as the error in line depth. 
We note that the `effective' S/N of RSGs could be under- or overestimated to some extent because, for example, weak stellar absorption lines that are not recognized in the chosen continuum regions may exist in reality. 
Moreover, the wavelength ranges in which many absorption features exist may be affected by the normalization error more than indicated by the `effective' S/N because it is harder to trace the continuum at these regions than the selected regions without absorption lines.

\section{Calibration of the Temperature Scale using Red Giants}\label{sec:ConstructingLDR}

In this section, we establish the key relations of this work, i.e., a set of the reliable empirical relations between the LDR and $T_{\mathrm{eff}}$, in the following strategy. 
First, we choose \ion{Fe}{i} lines that are relatively free from blending by other lines. 
Then we find the best pairs whose LDRs show a well-defined correlation with $T_{\mathrm{eff}}$. 

For the spectral synthesis employed in this section, we developed a wrapper software in \textsc{Python3} of the spectral synthesis code MOOG~\citep{Sneden1973,Sneden2012}\footnote{We used the February-2017 version of MOOG further modified by M. Jian~(\url{https://github.com/MingjieJian/moog_nosm}). }. 
MOOG synthesizes spectra, assuming the 1D LTE with the plane-parallel geometry. 
We remark that this assumption does not affect our final determination of $T_{\mathrm{eff}}$, which relies only on the empirical LDR--$T_{\mathrm{eff}}$ relations, though the selection of the candidate lines could be affected to some extent. 
The grid of the model atmospheres of RSGs and red giants was taken from the MARCS~\citep{Gustafsson2008} with the spherical geometry~($M=2M_{\odot }$ for most cases) assumed, and the model atmosphere at a set of finer-scale stellar parameters was obtained with linear interpolation. 
We note that although the geometry of the radiative transfer code~(plane parallel) is different from that of the model atmospheres assumed here~(spherical), the difference in their geometries is known to induce only a small effect in general on the synthesized spectra~\citep{Heiter2006}. 
As for the line list, we used the third release of the Vienna Atomic Line Database~\citep[VALD3;][]{Ryabchikova2015} as the main source. 
The molecular lines in the \textit{Y} and \textit{J} bands contained in the VALD3 are limited to \ce{CH}, \ce{OH}, \ce{SiH}, \ce{C2}, \ce{CN} and \ce{CO}, among which only \ce{CN} gives a significant absorption in the spectra of our target RSGs and red giants. 
In addition, we adopted the line list of \ce{FeH} by B. Plez~(private communication)\footnote{\url{https://www.lupm.in2p3.fr/users/plez/}}. 
As a result of the extensive examination, we found that lines of \ce{FeH} appeared in the spectra of the RSGs with a depth up to $\sim 10\%$ and that those lines were well reproduced by synthesized spectra with a dissociation energy of $1.59\ur{eV}$~\citep{Schultz1991}.

\subsection{Line selection}\label{ssec:LineSelection}

We chose the \ion{Fe}{i} lines that are sufficiently strong and relatively free from blending out of the list of the \ion{Fe}{i} lines in the VALD3. 
Since our aim is to estimate $T_{\mathrm{eff}}$ of RSGs using the LDR-$T_{\mathrm{eff}}$ relations calibrated with red giants in spite of the difference in the surface gravity between the two groups, it is necessary to choose the \ion{Fe}{i} lines that are least sensitive to the surface gravity. 
The LDR of a pair of lines from two different elements, especially those with different ionization energies, is known to show dependence on the surface gravity~\citepalias{Jian2020}. 
Also, depths of molecular lines are significantly different between RSGs and red giants because of their strong dependence on the surface gravity~\citep{Lancon2007}, non-solar CNO abundances~\citep{Lambert1984,Ekstrom2012} and existence of MOLspheres in RSGs~\citep{Tsuji2000,Kervella2009}. 
Therefore, we selected the \ion{Fe}{i} lines that are sufficiently free from blending with lines of molecules and atoms other than \ion{Fe}{i} itself. 
Moreover, contamination by other \ion{Fe}{i} lines may also introduce a gravity effect, given that the LDRs of the observed depths of blended lines are affected by microturbulent and macroturbulent velocities~\citep{Biazzo2007}, both of which tend to be larger at lower surface gravities~\citep{Holtzman2018}. 
Therefore, we also checked the contamination by other \ion{Fe}{i} lines. 

\begin{figure}
\centering 
\includegraphics[width=\columnwidth ]{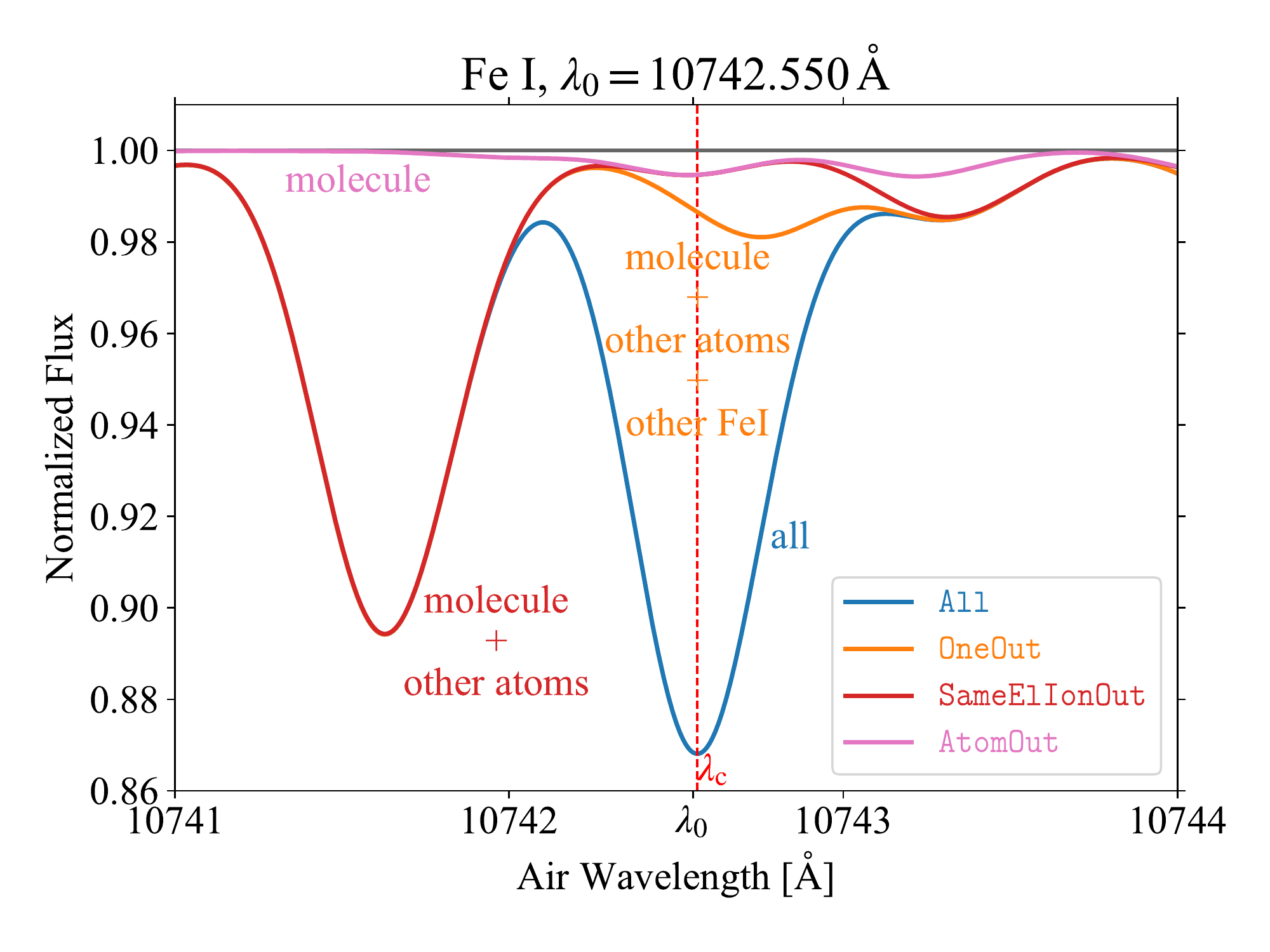}
\caption{Examples of the synthesized spectra of RSG3 around \ion{Fe}{i} $\lambda 10742.550\,\text{\AA }$~($\lambda _{0}$) with different lines included. Cyan, orange, red and pink lines show the spectra of \texttt{All}, \texttt{OneOut}, \texttt{SameElIonOut} and \texttt{AtomOut}, respectively, defined in the main text. The peak wavelength in \texttt{All}~($\lambda _{\mathrm{c}}$) was indicated with the vertical orange dashed line. We note that this line in this figure shows one of the most complex blended lines among the final set of the selected lines. }
\label{fig:EWcompLDR}
\end{figure}

In order to evaluate the effect of the potential blending, we used synthetic spectra of the two model RSGs~(RSG1 and RSG2 in \autoref{table:ImaginaryRSGAtmos}) and those of the five coolest red giants~(\textmu \ Leo, Alphard, Aldebaran, \textalpha \ Cet and \textdelta \ Oph). 
The stellar parameters of RSG1 and RSG2 roughly correspond to those of the warmest and coolest RSGs in our sample, respectively, and the line blendings in RSG2 are expected be severer than those in all the sample observed RSGs. 
With each set of stellar parameters we measured four types of depths~($d_{\mathtt{All}}$, $d_{\mathtt{OneOut}}$, $d_{\mathtt{SameElIonOut}}$ and $d_{\mathtt{AtomOut}}$) for each \ion{Fe}{i} line in VALD3~(let us use $\lambda _{0}$ to denote the centre wavelength listed in VALD3) as follows. 
First, we synthesized the respective four types of spectra around the target line with different groups of the lines included: (1)~\texttt{All}---all the atomic and molecular lines, (2)~\texttt{OneOut}---all the lines except for the target \ion{Fe}{i} line, (3)~\texttt{SameElIonOut}---all the lines except any \ion{Fe}{i} lines and (4)~\texttt{AtomOut}---only the molecular lines~(see an example in \autoref{fig:EWcompLDR}). 
Next, we identified the absorption around $\lambda _{0}$ in the synthesized spectrum \texttt{All} and determined the peak wavelength, $\lambda _{\mathrm{c}}$, the value of which is different from $\lambda _{0}$ if the target line is blended, or identical to it if not. 
Finally, the respective depths from unity at $\lambda _{\mathrm{c}}$ in all the four types of the synthesized spectra were measured. 
The reason why we measured the depths at $\lambda _{\mathrm{c}}$ rather than $\lambda _{0}$ in the synthesized spectra is that we measured the depth of each line in the observed spectra at the wavelength of the peak position around the line rather than the wavelength in the line list $\lambda _{0}$~(\autoref{ssec:LinePairSelection}). 

With these depths, we considered two types of criteria for selecting lines. 
First, we imposed $d_{\mathtt{All}}>0.02$ on the synthetic spectra of RSG1. 
When two or more lines were assigned to the same wavelength of minimum $\lambda _{\mathrm{c}}$ in RSG1, we would consider only the line that contributes most to the peak, i.e. the line with the smallest $d_{\mathtt{OneOut}}$, as a candidate. 
Then, we imposed the following three criteria, which should be satisfied in all the seven synthetic spectra of the RSGs and the red giants: 
\begin{equation}
\begin{cases}
d_{\mathtt{OneOut}}/d_{\mathtt{All}}<0.5 \\
d_{\mathtt{SameElIonOut}}/d_{\mathtt{All}}<0.3 \\
d_{\mathtt{AtomOut}}/d_{\mathtt{All}}<0.2
\end{cases}\text{.}
\end{equation}
Applying the combination of all the above-mentioned criteria to the current line list left $76$ \ion{Fe}{i} lines in total~($41$ in the \textit{Y} band and $35$ in \textit{J}). 
We note that the index $d_{\mathtt{OneOut}}/d_{\mathtt{All}}$ had been used in evaluating line blending in some of our recent papers~(\citealt{Kondo2019}; \citealt{Matsunaga2020}; \citetalias{Jian2020}); the difference in this work was that two slightly different indices, $d_{\mathtt{SameElIonOut}}$ and $d_{\mathtt{AtomOut}}$, were added to the condition to give tighter constraints. 

Furthermore, we considered the following six points, with which some candidate lines would be filtered out. 
First, some of the observed lines were found to be contaminated by unexpected stellar lines. 
For example, though \ion{Fe}{i} $\lambda 12393.067\,\text{\AA }$ is expected to be blended only with \ion{Fe}{i} $\lambda 12393.281\,\text{\AA }$ in the synthetic spectrum and to meet the criteria described above, the peak wavelengths $\lambda _{c}$ in the observed spectra of all the red giants were found to be systematically shorter by $\sim 0.25\,\text{\AA }$ than the wavelength~($\lambda _{0}$) listed in the VALD3 line list; this fact implies that this line is blended with an unknown line(s) with its peak wavelength at around $12392.7\,\text{\AA }$. 
Second, some of the observed lines were contaminated by residuals from telluric subtraction or suffered from imperfect continuum normalization due to many stellar lines, especially \ce{CN} molecular lines, concentrated in the close vicinity of the line. 
Third, some of the observed lines were found to be much weaker than those in the synthetic ones possibly due to the inaccurate oscillator strength values in the VALD3 line list. 
For example, the depths of \ion{Fe}{i} $\lambda 10462.155\,\text{\AA }$ were $0.016\text{--}0.028$ in the synthesized spectra of the four coolest red giants and this line meets the criteria, but the observed depths are too small, $0.010\text{--}0.019$. 
Fourth, we excluded two lines, \ion{Fe}{i} $\lambda 10780.694$ and $\lambda 12340.481\,\text{\AA }$, because they were affected with diffuse interstellar bands $\lambda 10780$ and $\lambda 12337$, respectively, reported by \citet{Hamano2015}. 
Fifth, hydrogen Paschen series and helium $10830\,\text{\AA }$ lines are known to show large variations between stars, especially in supergiant stars~\citep[e.g.][]{Obrien1986,Huang2012}, and hence may affect the blending significantly. However, we found that none of the selected lines were affected by them in our case. 
Sixth, \citet{Matsunaga2020} detected dozens of lines that are not found in the available line lists in the \textit{YJ}-band spectra of supergiants~($4000<T_{\mathrm{eff}}<7200\ur{K}$), and these lines may affect our result significantly. However, all the lines selected here were separated by more than $30\ur{\si{km.s^{-1}}}$ from these uncatalogued lines~(Table 3 of \citealt{Matsunaga2020}), and thus we safely ignored this potential influence. 

Applying all these points to the candidate lines, we obtained in the end $52$ \ion{Fe}{i} lines in total~($28$ in the \textit{Y} band and $24$ in \textit{J}) and summarize the result in \autoref{table:LDRFeLines}. 
We note that \citetalias{Taniguchi2018} considered another criterion requiring that the depth of each line be smaller than $0.5$ in the observed spectra of Arcturus; this condition is satisfied in all the lines that we selected. 

Then, we measured the depths of the selected $52$ \ion{Fe}{i} lines in each observed spectrum by fitting a quadratic function to three~(or four) pixels centred at the peak of each line~\citep{Strassmeier2000}. 
Here we define $d_{i}^{(n)}$ as the line depth from the continuum level to the peak of the fitted function, where $i$ denotes the ID number of the line and $(n)$ does the ID number of the star. 
We ignored the measurements if $d_{i}^{(n)}<0.02$, if the continuum around the line was not well normalized~(only $\lambda 9889.0351\,\text{\AA }$ in Aldebaran) or if the measured centre wavelength $\lambda _{\mathrm{c}}$ was not within $\pm 10\ur{\si{km.s^{-1}}}$ of the corresponding wavelength $\lambda_{0}$ in the VALD3 line list. 
We considered the reciprocal of the `effective' S/N per pixel estimated in \autoref{sssec:SNEstimateReal} to be the error in $d_{i}^{(n)}$, which includes both the systematic errors~(e.g., in the telluric subtraction and the continuum normalization) and the statistical errors~(Poisson, read-out and background noises). 

\begin{table}
\centering 
\caption{List of the 52 selected \ion{Fe}{i} lines. The last column is defined when the line is used in the final set of the LDR--$T_{\mathrm{eff}}$ relations, and shows the ID number of the LDR. }
\label{table:LDRFeLines}
\begin{tabular}{c cl cl cc} \toprule 
 & \multicolumn{2}{c}{Lower-level} & \multicolumn{2}{c}{Upper-level} & $\log gf$ & LDR \\
$\lambda _{\mathrm{air}}$ [\AA ] & $E$ [eV] & Term & $E$ [eV] & Term & [dex] & ID \\ \midrule 
$9800.3075$ & $5.0856$ & $x{}^{5}\mathrm{F}^{\circ }$ & $6.3503$ & $e{}^{5}\mathrm{G}$ & $-0.453$ &  \\
$9811.5041$ & $5.0117$ & $y{}^{7}\mathrm{P}^{\circ }$ & $6.2750$ & $e{}^{7}\mathrm{P}$ & $-1.362$ &  \\
$9868.1857$ & $5.0856$ & $x{}^{5}\mathrm{F}^{\circ }$ & $6.3416$ & $e{}^{7}\mathrm{F}$ & $-0.979$ & (1) \\
$9889.0351$ & $5.0331$ & $x{}^{5}\mathrm{F}^{\circ }$ & $6.2865$ & $e{}^{5}\mathrm{G}$ & $-0.446$ &  \\
$10065.045$ & $4.8349$ & $y{}^{3}\mathrm{D}^{\circ }$ & $6.0663$ & $e{}^{3}\mathrm{F}$ & $-0.289$ & (2) \\
$10081.393$ & $2.4242$ & $a{}^{3}\mathrm{P}_{2}$ & $3.6537$ & $z{}^{5}\mathrm{P}^{\circ }$ & $-4.537$ & (1) \\
$10155.162$ & $2.1759$ & $a{}^{5}\mathrm{P}$ & $3.3965$ & $z{}^{5}\mathrm{F}^{\circ }$ & $-4.226$ &  \\
$10216.313$ & $4.7331$ & $y{}^{3}\mathrm{D}^{\circ }$ & $5.9464$ & $e{}^{3}\mathrm{F}$ & $-0.063$ & (3) \\
$10218.408$ & $3.0713$ & $c{}^{3}\mathrm{P}$ & $4.2843$ & $z{}^{3}\mathrm{P}^{\circ }$ & $-2.760$ &  \\
$10265.217$ & $2.2227$ & $a{}^{5}\mathrm{P}$ & $3.4302$ & $z{}^{5}\mathrm{F}^{\circ }$ & $-4.537$ & (4) \\
$10332.327$ & $3.6352$ & $b{}^{3}\mathrm{D}$ & $4.8349$ & $y{}^{3}\mathrm{D}^{\circ }$ & $-2.938$ & (5) \\
$10340.885$ & $2.1979$ & $a{}^{5}\mathrm{P}$ & $3.3965$ & $z{}^{5}\mathrm{F}^{\circ }$ & $-3.577$ &  \\
$10347.965$ & $5.3933$ & $w{}^{5}\mathrm{D}^{\circ }$ & $6.5911$ & $f{}^{5}\mathrm{P}$ & $-0.551$ & (6) \\
$10353.804$ & $5.3933$ & $w{}^{5}\mathrm{D}^{\circ }$ & $6.5904$ & $h{}^{5}\mathrm{D}$ & $-0.819$ & (7) \\
$10364.062$ & $5.4457$ & $w{}^{5}\mathrm{D}^{\circ }$ & $6.6417$ & $f{}^{5}\mathrm{P}$ & $-0.960$ & (5) \\
$10395.794$ & $2.1759$ & $a{}^{5}\mathrm{P}$ & $3.3683$ & $z{}^{5}\mathrm{F}^{\circ }$ & $-3.393$ & (8) \\
$10423.027$ & $2.6924$ & $a{}^{3}\mathrm{G}$ & $3.8816$ & $z{}^{3}\mathrm{F}^{\circ }$ & $-3.616$ & (6) \\
$10423.743$ & $3.0713$ & $c{}^{3}\mathrm{P}$ & $4.2605$ & $z{}^{3}\mathrm{P}^{\circ }$ & $-2.918$ & (3) \\
$10532.234$ & $3.9286$ & $z{}^{3}\mathrm{D}^{\circ }$ & $5.1055$ & ${}^{3}\mathrm{P}$ & $-1.480$ & (8) \\
$10555.649$ & $5.4457$ & $w{}^{5}\mathrm{D}^{\circ }$ & $6.6200$ & $g{}^{5}\mathrm{F}$ & $-1.108$ & (9) \\
$10616.721$ & $3.2671$ & $b{}^{3}\mathrm{H}$ & $4.4346$ & $z{}^{3}\mathrm{G}^{\circ }$ & $-3.127$ & (4) \\
$10725.185$ & $3.6398$ & $b{}^{3}\mathrm{D}$ & $4.7955$ & $y{}^{3}\mathrm{D}^{\circ }$ & $-2.763$ & (7) \\
$10742.550$ & $3.6416$ & $b{}^{3}\mathrm{D}$ & $4.7955$ & $y{}^{3}\mathrm{D}^{\circ }$ & $-3.629$ & (2) \\
$10753.004$ & $3.9597$ & $z{}^{3}\mathrm{D}^{\circ }$ & $5.1124$ & ${}^{3}\mathrm{P}$ & $-1.845$ &  \\
$10754.753$ & $2.8316$ & $b{}^{3}\mathrm{P}$ & $3.9841$ & $z{}^{3}\mathrm{F}^{\circ }$ & $-4.523$ & (9) \\
$10783.050$ & $3.1110$ & $c{}^{3}\mathrm{P}$ & $4.2605$ & $z{}^{3}\mathrm{P}^{\circ }$ & $-2.567$ &  \\
$10818.274$ & $3.9597$ & $z{}^{3}\mathrm{D}^{\circ }$ & $5.1055$ & ${}^{3}\mathrm{P}$ & $-1.948$ & (10) \\
$10849.465$ & $5.5392$ & $e{}^{5}\mathrm{D}$ & $6.6817$ & ${}^{5}\mathrm{D}^{\circ }$ & $-1.444$ & (10) \\
$11638.260$ & $2.1759$ & $a{}^{5}\mathrm{P}$ & $3.2410$ & $z{}^{5}\mathrm{D}^{\circ }$ & $-2.214$ &  \\
$11681.594$ & $3.5465$ & $a{}^{1}\mathrm{D}$ & $4.6076$ & $y{}^{3}\mathrm{F}^{\circ }$ & $-3.615$ &  \\
$11783.265$ & $2.8316$ & $b{}^{3}\mathrm{P}$ & $3.8835$ & $z{}^{3}\mathrm{D}^{\circ }$ & $-1.574$ &  \\
$11882.844$ & $2.1979$ & $a{}^{5}\mathrm{P}$ & $3.2410$ & $z{}^{5}\mathrm{D}^{\circ }$ & $-1.668$ &  \\
$11884.083$ & $2.2227$ & $a{}^{5}\mathrm{P}$ & $3.2657$ & $z{}^{5}\mathrm{D}^{\circ }$ & $-2.083$ &  \\
$12119.494$ & $4.5931$ & $d{}^{3}\mathrm{F}$ & $5.6158$ & $y{}^{3}\mathrm{G}^{\circ }$ & $-1.635$ &  \\
$12190.098$ & $3.6352$ & $b{}^{3}\mathrm{D}$ & $4.6520$ & $y{}^{3}\mathrm{F}^{\circ }$ & $-2.330$ &  \\
$12213.336$ & $4.6382$ & $y{}^{5}\mathrm{P}^{\circ }$ & $5.6531$ & $e{}^{5}\mathrm{D}$ & $-1.845$ &  \\
$12267.888$ & $3.2740$ & $a{}^{3}\mathrm{D}$ & $4.2843$ & $z{}^{3}\mathrm{P}^{\circ }$ & $-4.368$ & (11) \\
$12556.996$ & $2.2786$ & $a{}^{3}\mathrm{P}_{2}$ & $3.2657$ & $z{}^{5}\mathrm{D}^{\circ }$ & $-3.626$ & (12) \\
$12615.928$ & $4.6382$ & $y{}^{5}\mathrm{P}^{\circ }$ & $5.6207$ & $e{}^{5}\mathrm{D}$ & $-1.517$ & (11) \\
$12638.703$ & $4.5585$ & $y{}^{5}\mathrm{P}^{\circ }$ & $5.5392$ & $e{}^{5}\mathrm{D}$ & $-0.783$ &  \\
$12648.741$ & $4.6070$ & $y{}^{5}\mathrm{P}^{\circ }$ & $5.5869$ & $e{}^{5}\mathrm{D}$ & $-1.140$ & (12) \\
$12789.450$ & $5.0095$ & $x{}^{5}\mathrm{D}^{\circ }$ & $5.9787$ & $e{}^{5}\mathrm{F}$ & $-1.514$ &  \\
$12807.152$ & $3.6398$ & $b{}^{3}\mathrm{D}$ & $4.6076$ & $y{}^{3}\mathrm{F}^{\circ }$ & $-2.452$ &  \\
$12808.243$ & $4.9880$ & $x{}^{5}\mathrm{D}^{\circ }$ & $5.9558$ & $e{}^{5}\mathrm{F}$ & $-1.362$ &  \\
$12824.859$ & $3.0176$ & $b{}^{3}\mathrm{G}$ & $3.9841$ & $z{}^{3}\mathrm{F}^{\circ }$ & $-3.835$ &  \\
$12879.766$ & $2.2786$ & $a{}^{3}\mathrm{P}_{2}$ & $3.2410$ & $z{}^{5}\mathrm{D}^{\circ }$ & $-3.458$ &  \\
$12896.118$ & $4.9130$ & $x{}^{5}\mathrm{D}^{\circ }$ & $5.8741$ & $e{}^{5}\mathrm{F}$ & $-1.424$ &  \\
$12934.666$ & $5.3933$ & $w{}^{5}\mathrm{D}^{\circ }$ & $6.3515$ & $e{}^{7}\mathrm{G}$ & $-0.948$ &  \\
$13006.684$ & $2.9904$ & $b{}^{3}\mathrm{G}$ & $3.9433$ & $z{}^{3}\mathrm{F}^{\circ }$ & $-3.744$ &  \\
$13014.841$ & $5.4457$ & $w{}^{5}\mathrm{D}^{\circ }$ & $6.3981$ & $f{}^{5}\mathrm{F}$ & $-1.693$ &  \\
$13098.876$ & $5.0095$ & $x{}^{5}\mathrm{D}^{\circ }$ & $5.9558$ & $e{}^{5}\mathrm{F}$ & $-1.290$ &  \\
$13147.920$ & $5.3933$ & $w{}^{5}\mathrm{D}^{\circ }$ & $6.3360$ & $f{}^{5}\mathrm{F}$ & $-0.814$ &  \\ \bottomrule 
\end{tabular}
\end{table}

\subsection{Line-pair selection: method}\label{ssec:LinePairSelection}

We searched for the best set of line pairs of which the LDR--$T_{\mathrm{eff}}$ relations yield the most precise estimates of $T_{\mathrm{eff}}$. 
Here, we followed the procedures in \citetalias{Taniguchi2018} and \citetalias{Jian2020} except that we combined the lists of \ion{Fe}{i} lines in different echelle orders in each of the \textit{Y} and \textit{J} bands. 
The basic idea of the process is to select the set of line pairs that meet the following conditions best: (1)~high precision in reproducing $T_{\mathrm{eff}}$ of our sample stars and (2)~small difference in wavelength between the two lines of each line pair.
The latter condition is desirable in the sense that different instruments from WINERED in future observations would have a better chance to be capable of detecting both lines of the line pair simultaneously. 

First, we evaluated the relation between LDR and $T_{\mathrm{eff}}$ of all the possible line pairs individually. 
In this case, $211$ and $128$ pairs in the \textit{Y} and \textit{J} bands, respectively, were found to have two lines that were both detected in more than four stars and have excitation potentials~(EPs) separated by more than $1\ur{eV}$. 
For each pair, denoted with the subscript $j$, of all the $211+128$ line pairs, we calculated the LDR denoted as $r_{j}^{(n)}$ and its error. 
We plotted $T_{\mathrm{eff}}$ against the common logarithms of the LDRs~($\log r_{j}$) for each pair and determined the regression line $T_{\mathrm{eff}}=a_{j}\log r_{j}+b_{j}$, using the Weighted Total Least Squares method~\citep[see][for a review]{Markovsky2007}.
For this regression, the weight of each point was given by 
\begin{equation}
w_{j}^{(n)}=\left[\left(\Delta T_{\mathrm{eff}}^{(n)}\right)^{2}+{a_{j}}^{2}\left(\sigma _{j}^{(n)}\right)^{2}\right]^{-1}\text{,}
\end{equation}
where $\Delta T_{\mathrm{eff}}^{(n)}$ and $\sigma _{j}^{(n)}$ indicate the standard errors of $T_{\mathrm{eff}}^{(n)}$ in literature and of $\log r_{j}^{(n)}$ for each star, respectively.
The weighted dispersion with regard to each regression line was then calculated as
\begin{equation}\label{eq:sigma}
\sigma _{j}=\sqrt{\frac{N}{N-2}\frac{\displaystyle \sum _{n=1}^{N}w_{j}^{(n)}\left[T_{\mathrm{eff}}^{(n)}-(a_{j}\log r_{j}^{(n)}+b_{j})\right]^{2}}{\displaystyle \sum _{n=1}^{N}w_{j}^{(n)}}}\text{.}
\end{equation}
We set the threshold of $\sigma _{j}<150\ur{K}$ for the pair to be valid. 
As a result, $41$ and $6$ line pairs in the \textit{Y} and \textit{J} bands, respectively, were left in the subsequent analyses.
We also set another threshold $a_{j}>0$, but no line pairs were rejected with this condition. 

Second, we searched for the optimal set of the line pairs for each of the \textit{Y} and \textit{J} bands as follows. 
Each set of line pairs can be considered as an undirected graph whose nodes correspond to absorption lines and whose edges connecting the nodes correspond to line pairs.
We require that each line be used only once in each set, i.e. no duplication of a line in separate pairs are allowed.
This ensures that the errors in $r_{j}^{(n)}$ values in different LDR--$T_{\mathrm{eff}}$ relation be independent of each other and thus makes it straightforward to calculate the statistical errors of the combined $T_{\mathrm{eff}}$~(\autoref{eq:DeltaTLDR1} and \autoref{eq:DeltaTLDR2}). 
We determined the optimal matching, $M$, that meets our requirements as follows. 
We considered the maximum matching, where the number of the edges of the matching is as large as possible under the condition where no nodes are connected with more than one edge.
In the ideal case, the size of the maximum matching, $\abs{M}$, corresponds to a half of the number of all the selected lines. 
However, it did not in our case because many edges were rejected due to, for example, large scatters around the corresponding LDR--$T_{\mathrm{eff}}$ relations.
For a given maximum matching, $M_{k}$, of this undirected graph, we calculated the weighted mean temperature $T_{M_{k}}^{(n)}$ on the basis of the LDR--$T_{\mathrm{eff}}$ relations for each star. 
We also considered the difference in wavelength between the two lines of each line pair $j_{k,m}$, denoted as $\Delta \lambda _{m,k}$, and calculated the evaluation function $E(M_{k};e)$ defined as 
\begin{align}\label{eq:evalfunc}
E(M_{k};e)&=\sqrt{\frac{1}{N}\sum _{n=1}^{N}\left(T_{M_{k}}^{(n)}-T_{\mathrm{eff}}^{(n)}\right)^{2}}+e\sqrt{\frac{1}{\abs{M_{k}}}\sum _{m=1}^{\abs{M_{k}}}\left(\Delta \lambda _{k,m}\right)^{2}} \\
&=E_{T}(M_{k})+eE_{\lambda }(M_{k})\notag \text{,}
\end{align}
where $E_{T}(M_{k})$ represents the size of the error in redetermining $T_{\mathrm{eff}}$ of the nine stars for a given matching $M_{k}$, and $E_{\lambda }(M_{k})$ represents the wavelength difference of the line pairs and works as a penalty term.
There are different allowed combinations of the line pairs which form different maximum matchings, and for a given $e$ value we selected the one that gives the least $E(M_{k};e)$ as the optimal matching, $M_{k(e)}$.
We would determine the coefficient $e$ by considering how $E_{T}(M_{k(e)};e)$ and $E_{\lambda }(M_{k(e)};e)$ depend on $e$, as explained in the following subsection.

\subsection{Line-pair selection: results}\label{ssec:ResLDRSelection}

\begin{figure}
\centering 
\includegraphics[width=\columnwidth ]{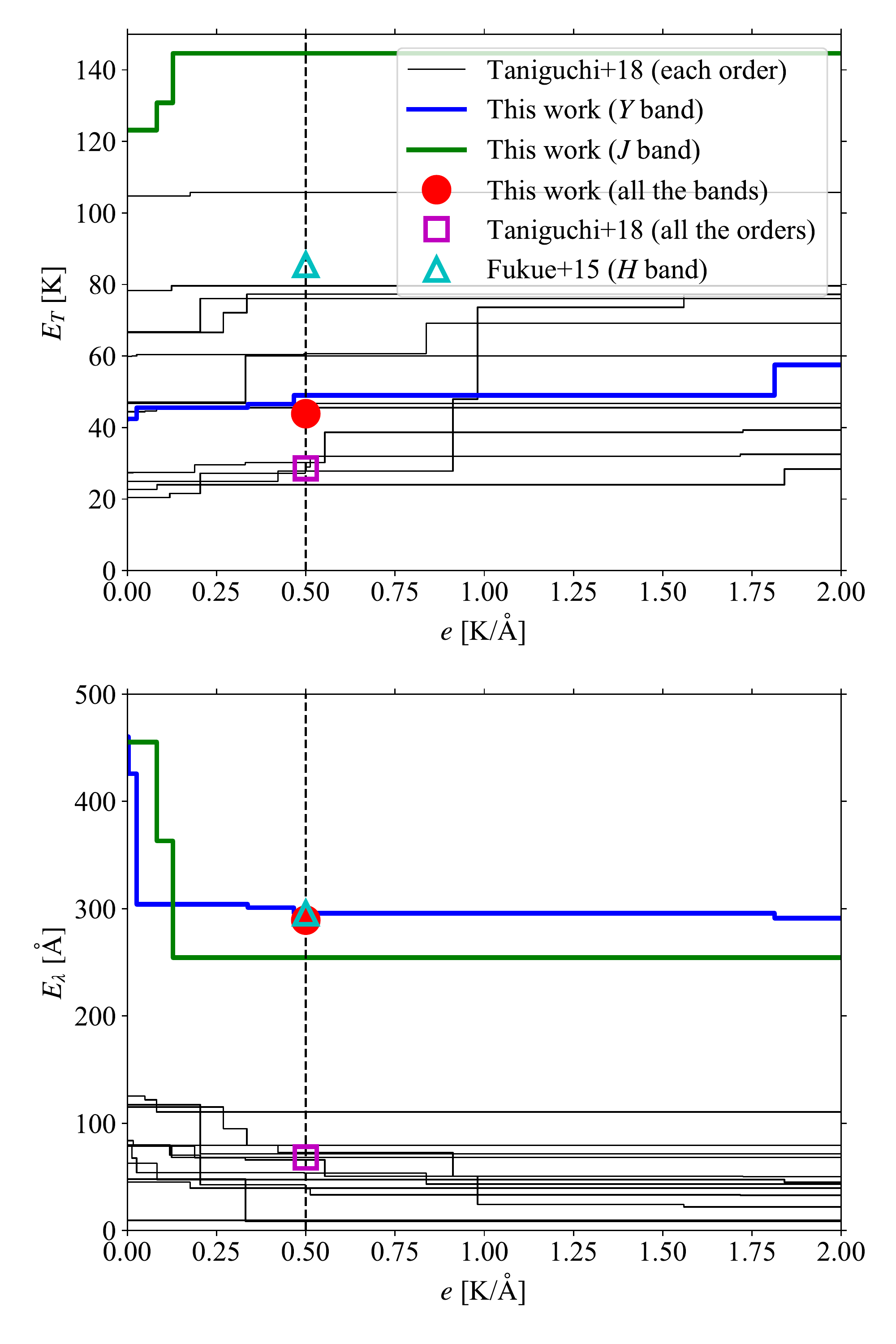}
\caption{Evaluation functions, $E_{T}$ and $E_{\lambda }$, of the coefficient $e$~(see text). Blue and green thick lines show those for the \textit{Y} and \textit{J} bands, respectively, in this work. Black thin lines show those of 57th--52nd and 48th--43rd orders by \citetalias{Taniguchi2018}. Red filled circles, magenta open squares and cyan open triangles show the evaluation functions of all the orders (or bands) combined, where the line pairs selected in this work, \citetalias{Taniguchi2018} and \citet{Fukue2015}, respectively, are employed. }
\label{fig:EvalPlot}
\end{figure}

We applied the procedure described in the previous subsection to our $28$ and $24$ nodes~(i.e. lines) and $41$ and $6$ edges~(i.e. line pairs) in the \textit{Y} and \textit{J} bands, respectively. 
We searched for the optimized matching, $M_{k(e)}$, that gives the smallest $E$ at each $e$ value between $0$ and $2\ur{K/\text{\AA }}$ to see how $e$ would affect the solutions.
\autoref{fig:EvalPlot} plots the values of $E_{T}$ and $E_{\lambda }$ for $M_{k(e)}$ with varying $e$. 
Larger the parameter $e$\ is, larger the weight on $E_{\lambda }$ is relative to $E_{T}$ in the total evaluation function $E$, and then the optimal matching and the values of $E_{T}$ and $E_{\lambda }$ change.
We decided to employ the same value, $e=0.5\ur{K/\text{\AA }}$, as \citetalias{Taniguchi2018} because $E_{T}$ at $e=0.5\ur{K/\text{\AA }}$ is similar to that at $e=0\ur{K/\text{\AA }}$, which optimizes the precision in the redetermined temperature, whereas $E_{\lambda }$ is sufficiently small. 
Moreover, the same matching would have been selected if another value of $e$ among a wide range of $e$ values~($0.39<e<1.81\ur{K/\text{\AA }}$) was employed. 

\begin{figure*}
\centering 
\includegraphics[width=\textwidth ]{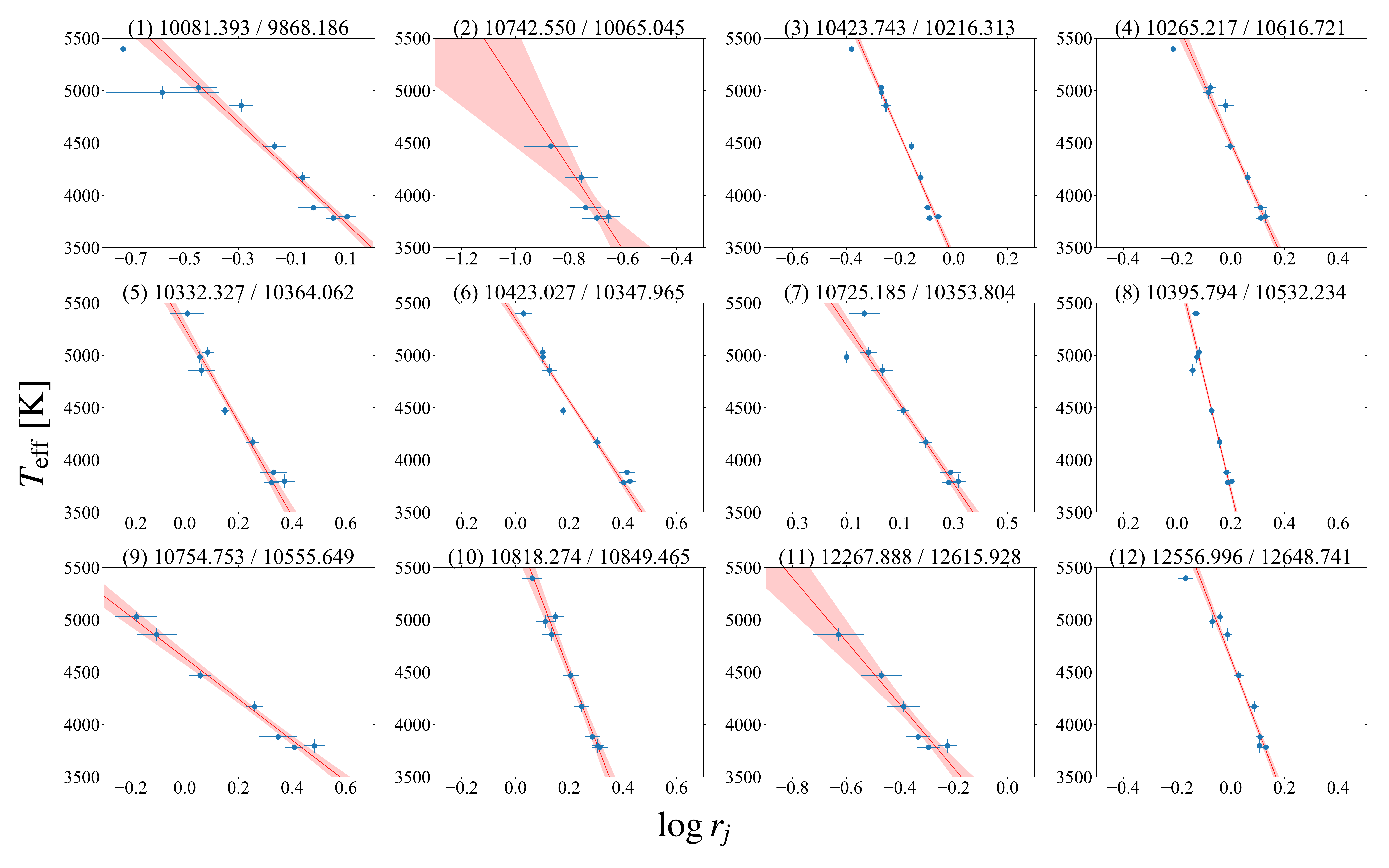}
\caption{LDR--$T_{\mathrm{eff}}$ relations of the twelve \ion{Fe}{i}--\ion{Fe}{i} line pairs that we selected~(also see \autoref{table:LDRFePairs}). The line-pair ID together with the wavelengths (\AA ) of low- and high- excitation lines are indicated at the top of each panel. Blue data points show the relation between the observed $\log (\mathrm{LDR})$ and effective temperatures in literature for red giants. Red lines indicate the best-fitting relations that we obtained, $T_{\mathrm{eff}}=a\log (\mathrm{LDR})+b$, with light-red shaded areas for the $1\sigma $ confidence intervals. }
\label{fig:LDRFePairs}
\end{figure*}

Consequently, we obtained $10$ and $2$ line pairs in the \textit{Y} and \textit{J} bands, respectively, i.e., $12$ line pairs in total~(\autoref{table:LDRFePairs} and \autoref{fig:LDRFePairs}). 
This set of line pairs made use of $24$ unique lines in the \textit{Y} and \textit{J} bands combined~(nb., no duplications of the lines are allowed; see \autoref{ssec:LinePairSelection}). 
This condition has only a weak effect on the statistical error, which could have been better by only $\sim 10\%$ even if we had used all the $41+6$ pre-selected pairs. 
The number of the selected lines~($52$ in total, or $28$ in \textit{Y} and $24$ in \textit{J}; see \autoref{ssec:LineSelection}) is about one-fourth of that used by \citetalias{Taniguchi2018}~($224$ in total, or $125$ in \textit{Y} and $99$ in \textit{J}), where atomic lines of various elements were used. 
Accordingly, the total number of the final line-pairs, $12$, is smaller than $81$ in \citetalias{Taniguchi2018}. 
Nevertheless, the number $12$ is larger than the number, $8$, of the \ion{Fe}{i} line pairs in \citetalias{Taniguchi2018}. 

The evaluation functions with $e=0.5\ur{K/\text{\AA }}$ with all the line pairs in the \textit{Y} and \textit{J} bands combined are $E_{T}^{\mathrm{all}}=43.9\ur{K}$ and $E_{\lambda }^{\mathrm{all}}=289.3\,\text{\AA }$. 
The former $E_{T}^{\mathrm{all}}$ in this work is only $1.5$ times larger than the counterpart in \citetalias{Taniguchi2018} despite the smaller number of the line pairs and smaller depths, both of which could have increased the statistical errors in $T_{M_{k}}^{(n)}$. 
The relatively small difference in $E_{T}^{\mathrm{all}}$ indicates that the errors in literature $T_{\mathrm{eff}}$~($19\text{--}65\ur{K}$) mainly contribute to $E_{T}^{\mathrm{all}}$ in \citetalias{Taniguchi2018}. 
In contrast, the latter $E_{\lambda }^{\mathrm{all}}$ in this work is $4.3$ times larger than that by \citetalias{Taniguchi2018} and is similar to that by \citet{Fukue2015} for the \textit{H} band. 
This is expected because \citet{Fukue2015} and we treated lines in each band together, whereas \citetalias{Taniguchi2018} treated those in each echelle order independently. 
These differences in $E_{T}$ and $E_{\lambda }$ are well demonstrated in \autoref{fig:EvalPlot}. 

\begin{table*}
\centering 
\caption{List of low- and high-excitation \ion{Fe}{i} lines and the LDR--$T_{\mathrm{eff}}$ relations; $a$ and $b$ denote the coefficients in $T_{\mathrm{eff}}=a\log r+b$, the three $S$ values~($S_{aa}$, $S_{ab}$ and $S_{bb}$) represent the variance-covariance matrix of $a$ and $b$, $N$ is the number of stars used in the fitting, and $\sigma $ is the weighted dispersion given by \autoref{eq:sigma}. The line pair (6) in this table is included in Table~4 of \citetalias{Taniguchi2018} with the ID~(22). }
\label{table:LDRFePairs}
\scalebox{0.97}{
\begin{tabular}{cc crr crr rrrrrrr}\toprule 
 & & \multicolumn{3}{c}{Low-excitation line} & \multicolumn{3}{c}{High-excitation line} & \multicolumn{7}{c}{LDR--$T_{\mathrm{eff}}$ relation} \\ \cmidrule(lr){3-5}\cmidrule(lr){6-8}\cmidrule(lr){9-15}
ID & Band & Order & \multicolumn{1}{c}{$\lambda _{\mathrm{air}}$ [\AA ]} & \multicolumn{1}{c}{EP} & Order & \multicolumn{1}{c}{$\lambda _{\mathrm{air}}$ [\AA ]} & \multicolumn{1}{c}{EP} & \multicolumn{1}{c}{$a$} & \multicolumn{1}{c}{$b$} & \multicolumn{1}{c}{$S_{aa}$} & \multicolumn{1}{c}{$S_{ab}$} & \multicolumn{1}{c}{$S_{bb}$} & $N$ & \multicolumn{1}{c}{$\sigma $} \\
 & & & & [eV] & & & [eV] & [K] & [K] & [$10^{3}\mathrm{K}^{2}$] & [$10^{2}\mathrm{K}^{2}$] & [$10^{2}\mathrm{K}^{2}$] & & [K]\\ \midrule 
(1) & \textit{Y} & m56 & $10081.393$ & $2.424$ & m57 & $9868.1857$ & $5.086$ & $-2408$ & $3976$ & $37.5$ & $26.0$ & $15.5$ & $9$ & $133$ \\
(2) & \textit{Y} & m52 & $10742.550$ & $3.642$ & m56 & $10065.045$ & $4.835$ & $-3891$ & $1149$ & $3803.8$ & $26919.2$ & $19154.3$ & $5$ & $131$ \\
(3) & \textit{Y} & m54 & $10423.743$ & $3.071$ & m55 & $10216.313$ & $4.733$ & $-5858$ & $3401$ & $88.7$ & $157.1$ & $35.1$ & $9$ & $121$ \\
(4) & \textit{Y} & m55 & $10265.217$ & $2.223$ & m53 & $10616.721$ & $3.267$ & $-5731$ & $4501$ & $214.3$ & $-80.9$ & $19.1$ & $9$ & $114$ \\
(5) & \textit{Y} & m54 & $10332.327$ & $3.635$ & m54 & $10364.062$ & $5.446$ & $-4492$ & $5259$ & $162.7$ & $-261.5$ & $59.2$ & $9$ & $119$ \\
(6) & \textit{Y} & m54 & $10423.027$ & $2.692$ & m54 & $10347.965$ & $5.393$ & $-3912$ & $5346$ & $42.7$ & $-91.5$ & $26.3$ & $9$ & $127$ \\
(7) & \textit{Y} & m52 & $10725.185$ & $3.640$ & m54 & $10353.804$ & $5.393$ & $-3791$ & $4910$ & $93.5$ & $-141.7$ & $38.8$ & $9$ & $137$ \\
(8) & \textit{Y} & m54 & $10395.794$ & $2.176$ & m53 & $10532.234$ & $3.929$ & $-10723$ & $5853$ & $438.5$ & $-624.0$ & $101.2$ & $9$ & $144$ \\
(9) & \textit{Y} & m52 & $10754.753$ & $2.832$ & m53 & $10555.649$ & $5.446$ & $-1966$ & $4636$ & $34.7$ & $-95.5$ & $40.6$ & $7$ & $74$ \\
(10) & \textit{Y} & m52 & $10818.274$ & $3.960$ & m52 & $10849.465$ & $5.539$ & $-6724$ & $5848$ & $720.5$ & $-1601.5$ & $405.4$ & $9$ & $84$ \\
(11) & \textit{J} & m46 & $12267.888$ & $3.274$ & m44 & $12615.928$ & $4.638$ & $-3025$ & $2980$ & $445.8$ & $1400.6$ & $482.9$ & $6$ & $123$ \\
(12) & \textit{J} & m45 & $12556.996$ & $2.279$ & m44 & $12648.741$ & $4.607$ & $-6714$ & $4629$ & $214.2$ & $-82.8$ & $16.6$ & $9$ & $134$ \\ \bottomrule 
\end{tabular}
}
\end{table*}

\subsection{Re-determination of the effective temperatures of red giants}\label{ssec:ReDetTeffRGB}

With each line pair $j$, the effective temperature and its error of a target star~(RSG in our case) would be estimated to be 
\begin{align}
&T_{j}=a_{j}\log r_{j}+b_{j}\text{,} \\
&\Delta T_{j}=\sqrt{(a_{j}\Delta \log r_{j})^{2}+S_{aa}(\log r_{j})^{2}+2S_{ab}\log r_{j}+S_{bb}}\text{,}
\end{align}
where $\begin{pmatrix}S_{aa} & S_{ab} \\ S_{ab} & S_{bb}\end{pmatrix}$ is the variance-covariance matrix of the coefficients $(a_{j},b_{j})$ of the regression line. 
The value referred to as the LDR effective temperature $T_{\mathrm{LDR}}$ of the target star is defined as the weighted mean of the temperatures estimated from the available line pairs among the  $12$ line pairs~($T_{j}\pm \Delta T_{j}$; for $j=1,\cdots ,N_{\mathrm{pair}}$).
We define two forms of the error for $T_{\mathrm{LDR}}$, given by the following two formulae, and calculated both of these for each target, 
\begin{equation}
\Delta T_{\mathrm{LDR}}^{\mathrm{Eq}\ref{eq:DeltaTLDR1}}=\left[\frac{1}{N_{\mathrm{pair}}-1}\frac{\displaystyle \sum _{j=1}^{N_{\mathrm{pair}}}\frac{(T_{j}-T_{\mathrm{LDR}})^{2}}{{\Delta T_{j}}^{2}}}{\displaystyle \sum _{j=1}^{N_{\mathrm{pair}}}\frac{1}{{\Delta T_{j}}^{2}}}\right]^{1/2}\label{eq:DeltaTLDR1}
\end{equation}
\begin{equation}
\Delta T_{\mathrm{LDR}}^{\mathrm{Eq}\ref{eq:DeltaTLDR2}}=\left[\sum _{j=1}^{N_{\mathrm{pair}}}\frac{1}{{\Delta T_{j}}^{2}}\right]^{-1/2}\text{.}\label{eq:DeltaTLDR2}
\end{equation}
The former~($\Delta T_{\mathrm{LDR}}^{\mathrm{Eq}\ref{eq:DeltaTLDR1}}$) is the weighted standard error of $T_{\mathrm{LDR}}$ used by \citetalias{Taniguchi2018}, and the latter~($\Delta T_{\mathrm{LDR}}^{\mathrm{Eq}\ref{eq:DeltaTLDR2}}$) is the error propagated from the errors in the LDRs and coefficients of the relations. 

\begin{table*}
\centering 
\caption{Effective temperatures of nine red giants in kelvin. $T_{\mathrm{eff}}$ is the literature effective temperatures adopted in this work. $\Delta T_{\mathrm{LDR}}^{\mathrm{Eq}\ref{eq:DeltaTLDR2}}$ in group (T18) is calculated in this work, whereas $T_{\mathrm{LDR}}$ and $\Delta T_{\mathrm{LDR}}^{\mathrm{Eq}\ref{eq:DeltaTLDR1}}$ in group (T18) are adopted from \citetalias{Taniguchi2018}, in which they calculated the values in the same way as in this work. }\label{table:CompareTLDRFe}
\begin{tabular}{lc cccc cccc}\toprule 
 & & \multicolumn{4}{c}{\citetalias{Taniguchi2018}~(T18)} & \multicolumn{4}{c}{This work~(TW)} \\ \cmidrule(lr){3-6}\cmidrule(lr){7-10}
Object & $T_{\mathrm{eff}}$ & $T_{\mathrm{LDR}}$ & $\Delta T_{\mathrm{LDR}}^{\mathrm{Eq}\ref{eq:DeltaTLDR1}}$ & $\Delta T_{\mathrm{LDR}}^{\mathrm{Eq}\ref{eq:DeltaTLDR2}}$ & $N_{\mathrm{pair}}$ & $T_{\mathrm{LDR}}$ & $\Delta T_{\mathrm{LDR}}^{\mathrm{Eq}\ref{eq:DeltaTLDR1}}$ & $\Delta T_{\mathrm{LDR}}^{\mathrm{Eq}\ref{eq:DeltaTLDR2}}$ & $N_{\mathrm{pair}}$ \\ \midrule 
\textepsilon \ Leo & $5398\pm 31$ & $5429$ & $24$ & $17$ & $42$ & $5420$ & $93$ & $60$ & $9$ \\
\textkappa \ Gem & $5029\pm 47$ & $4982$ & $13$ & $10$ & $60$ & $4953$ & $14$ & $32$ & $10$ \\
\textepsilon \ Vir & $4983\pm 61$ & $4996$ & $11$ & $10$ & $65$ & $4999$ & $26$ & $29$ & $9$ \\
Pollux & $4858\pm 60$ & $4829$ & $12$ & $11$ & $73$ & $4829$ & $50$ & $46$ & $11$ \\
\textmu \ Leo & $4470\pm 40$ & $4434$ & $15$ & $7$ & $80$ & $4517$ & $38$ & $26$ & $12$ \\
Alphard & $4171\pm 52$ & $4143$ & $9$ & $7$ & $79$ & $4134$ & $7$ & $26$ & $12$ \\
Aldebaran & $3882\pm 19$ & $3887$ & $6$ & $7$ & $78$ & $3905$ & $27$ & $42$ & $12$ \\
\textalpha \ Cet & $3796\pm 65$ & $3780$ & $9$ & $6$ & $79$ & $3737$ & $25$ & $32$ & $12$ \\
\textdelta \ Oph & $3783\pm 20$ & $3812$ & $4$ & $6$ & $79$ & $3833$ & $15$ & $29$ & $12$ \\ \bottomrule 
\end{tabular}
\end{table*}

\begin{figure*}
\centering 
\includegraphics[width=\textwidth ]{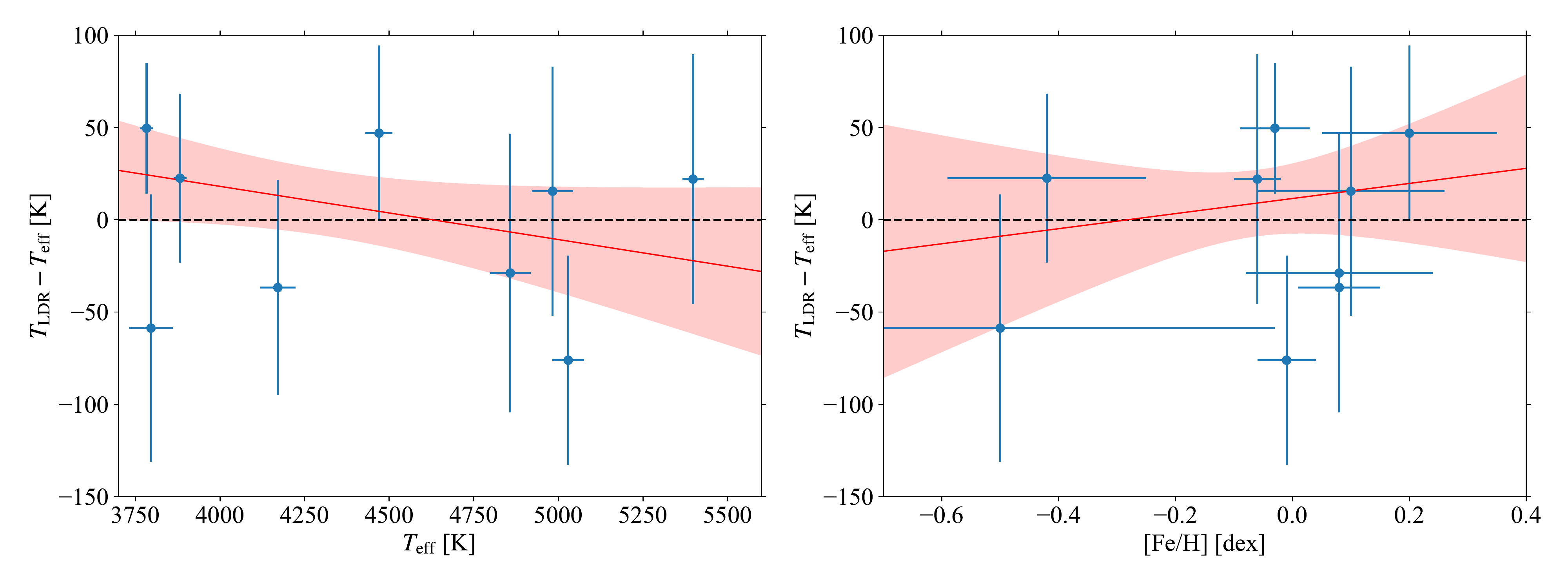}
\caption{Differences between the re-determined and literature effective temperatures~($T_{\mathrm{LDR}}$(TW) and $T_{\mathrm{eff}}$, respectively) for red giants are plotted against the literature stellar parameters~($T_{\mathrm{eff}}$ and [Fe/H]). Red lines show the linear regression lines of blue points with the light red shades showing the $1\sigma $ confidence intervals. }
\label{fig:CompareTLDRFe}
\end{figure*}

In order to examine the dependence of the relations on $T_{\mathrm{eff}}$ and [Fe/H], we determined $T_{\mathrm{LDR}}$ of the group of the nine red giants~(\autoref{table:ObsLog}).
\autoref{table:CompareTLDRFe} summarizes the re-determined effective temperatures and their standard errors in this work, together with those estimated on the basis of the relations in \citetalias{Taniguchi2018}. 
In the table and hereafter, the former and latter cases are abbreviated as `TW'~(as of This Work) and `T18', respectively, and the parameters based on either of them are distinguished with the corresponding abbreviated word in parentheses as the suffix; e.g., $\Delta T_{\mathrm{LDR}}^{\mathrm{Eq}\ref{eq:DeltaTLDR1}}$(TW) denotes the error of the LDR effective temperature on the basis of ~\autoref{eq:DeltaTLDR1} calculated with the relations in this work~(TW). 
To estimate the standard errors according to \autoref{eq:DeltaTLDR1} is, however, problematic in the case of the relations in this work for two reasons; (1)~the error of $\Delta T_{\mathrm{LDR}}^{\mathrm{Eq}\ref{eq:DeltaTLDR1}}$(TW) may be large because the number of the line pairs is small, and (2)~the dispersion of $T_{j}$, and hence $\Delta T_{\mathrm{LDR}}^{\mathrm{Eq}\ref{eq:DeltaTLDR1}}$(TW), may be underestimated because the nine red giants themselves were used to calibrate the LDR--$T_{\mathrm{eff}}$ relations. 
The errors according to \autoref{eq:DeltaTLDR2} are more likely to be closer to the true values as long as the errors of the LDRs are accurately estimated. 
In fact, $\Delta T_{\mathrm{LDR}}^{\mathrm{Eq}\ref{eq:DeltaTLDR1}}$(TW) tends to be significantly smaller than $\Delta T_{\mathrm{LDR}}^{\mathrm{Eq}\ref{eq:DeltaTLDR2}}$(TW) probably because of the aforementioned reasons. 
Therefore, we concluded that the latter, $\Delta T_{\mathrm{LDR}}^{\mathrm{Eq}\ref{eq:DeltaTLDR2}}$(TW), is more robust and reliable than $\Delta T_{\mathrm{LDR}}^{\mathrm{Eq}\ref{eq:DeltaTLDR1}}$(TW). 
In contrast, the error $\Delta T_{\mathrm{LDR}}^{\mathrm{Eq}\ref{eq:DeltaTLDR1}}$(T18) was found to be similar to $\Delta T_{\mathrm{LDR}}^{\mathrm{Eq}\ref{eq:DeltaTLDR2}}$(T18) probably because of the large number of line pairs, and thus we simply adopted $\Delta T_{\mathrm{LDR}}^{\mathrm{Eq}\ref{eq:DeltaTLDR1}}$(T18) as in \citetalias{Taniguchi2018}. 

\autoref{fig:CompareTLDRFe} demonstrates the differences between the re-determined effective temperatures $T_{\mathrm{LDR}}$(TW) and those in literature $T_{\mathrm{eff}}$. 
It shows no apparent correlation between the difference and either of $T_{\mathrm{eff}}$ and [Fe/H]. 
Our $T_{\mathrm{LDR}}$ was consistent with that in literature within $\sim 50\ur{K}$ over the entire range of $T_{\mathrm{eff}}$ and [Fe/H] of our sample red giants for the calibration. 
Besides, the error bars on the basis of $\Delta T_{\mathrm{LDR}}^{\mathrm{Eq}\ref{eq:DeltaTLDR2}}$(TW) and $\Delta T_{\mathrm{eff}}$ explain the scatter around zero well, indicating that the error estimates were reasonable.

\section{Determination of Effective Temperatures of Red Supergiants}\label{sec:TeffRSGs}

\subsection{LDR temperatures of red supergiants}\label{ssec:TLDRRSG}

\begin{table*}
\centering 
\caption{Effective temperatures of ten RSGs using different sets of LDR--$T_{\mathrm{eff}}$ relations in kelvin. $T_{\mathrm{TiO}}$(L05) shows the effective temperatures estimated with the optical \ce{TiO} bands by \citet{Levesque2005} for comparison. }
\label{table:TLDRRSG}
\begin{tabular}{lc cccc cccc}\toprule 
 & & \multicolumn{4}{c}{\citetalias{Taniguchi2018}~(T18)} & \multicolumn{4}{c}{This work~(TW)} \\ \cmidrule(lr){3-6}\cmidrule(lr){7-10}
Object & $T_{\mathrm{TiO}}$(L05) & $T_{\mathrm{LDR}}$ & $\Delta T_{\mathrm{LDR}}^{\mathrm{Eq}\ref{eq:DeltaTLDR1}}$ & $\Delta T_{\mathrm{LDR}}^{\mathrm{Eq}\ref{eq:DeltaTLDR2}}$ & $N_{\mathrm{pair}}$ & $T_{\mathrm{LDR}}$ & $\Delta T_{\mathrm{LDR}}^{\mathrm{Eq}\ref{eq:DeltaTLDR1}}$ & $\Delta T_{\mathrm{LDR}}^{\mathrm{Eq}\ref{eq:DeltaTLDR2}}$ & $N_{\mathrm{pair}}$ \\ \midrule 
\textzeta \ Cep & $4000$ & $4340$ & $22$ & $13$ & $72$ & $4077$ & $35$ & $53$ & $11$ \\
41 Gem & $3900$ & $4021$ & $16$ & $9$ & $81$ & $3944$ & $29$ & $45$ & $12$ \\
\textxi \ Cyg & $3800$ & $3989$ & $15$ & $10$ & $81$ & $3894$ & $25$ & $41$ & $12$ \\
V809 Cas & $3750$ & $3960$ & $18$ & $10$ & $77$ & $3772$ & $36$ & $47$ & $12$ \\
V424 Lac & $3800$ & $3934$ & $19$ & $13$ & $78$ & $3755$ & $49$ & $64$ & $12$ \\
\textpsi ${}^{1}$ Aur & $3750$ & $4049$ & $24$ & $10$ & $75$ & $3742$ & $65$ & $47$ & $12$ \\
TV Gem & $3700$ & $4008$ & $23$ & $13$ & $69$ & $3681$ & $109$ & $66$ & $9$ \\
BU Gem & $3800$ & $3970$ & $26$ & $14$ & $71$ & $3883$ & $62$ & $73$ & $10$ \\
Betelgeuse & $3650$ & $3862$ & $20$ & $13$ & $75$ & $3618$ & $40$ & $66$ & $12$ \\
NO Aur & $3700$ & $3849$ & $19$ & $11$ & $77$ & $3659$ & $34$ & $58$ & $12$ \\
\bottomrule 
\end{tabular}
\end{table*}

\begin{figure*}
\centering 
\includegraphics[width=\textwidth ]{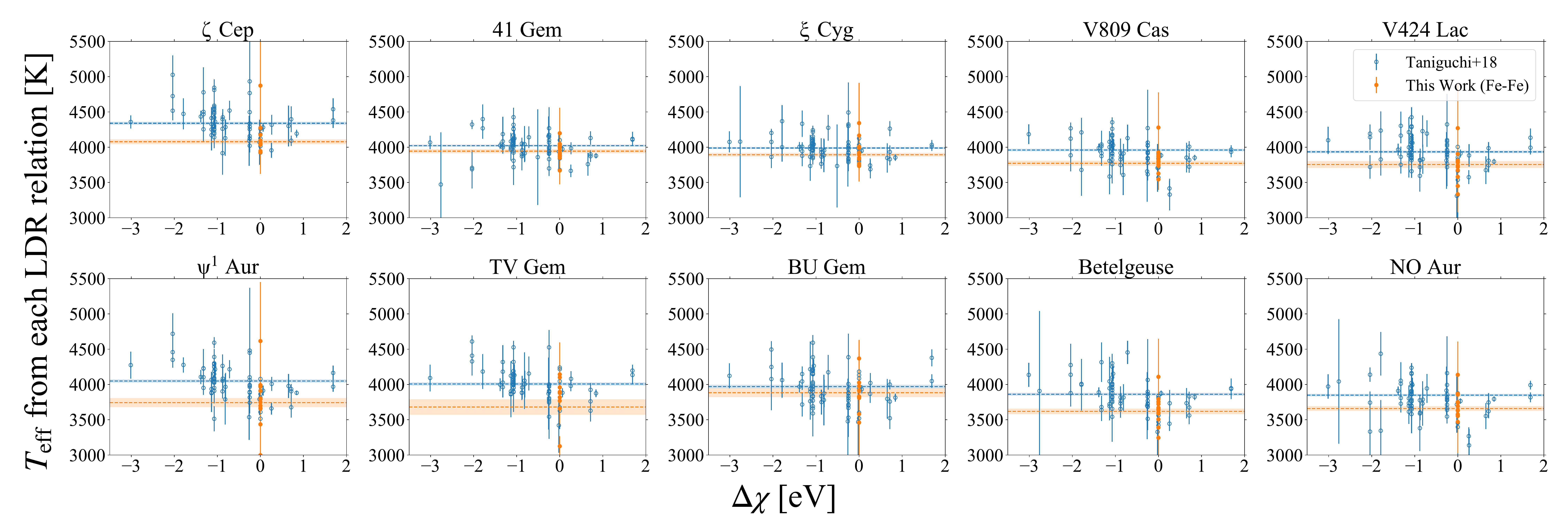}
\caption{Effective temperatures derived with the LDR relations~(blue) in \citetalias{Taniguchi2018} and~(orange) of this work for RSGs are plotted against the difference in the ionization energies of the two lines in each line pair. The weighted mean values $T_{\mathrm{LDR}}$ of the data points from the two sets of the relations are indicated by horizontal dashed lines in the respective colours, shaded with the width corresponding to the standard errors $\Delta T_{\mathrm{LDR}}^{\mathrm{Eq}\ref{eq:DeltaTLDR1}}$~(\autoref{eq:DeltaTLDR1}). }
\label{fig:IPDep}
\end{figure*}

Our new LDR--$T_{\mathrm{eff}}$ relations for \ion{Fe}{i}--\ion{Fe}{i} line pairs are supposed to be less affected by the surface gravity and line broadening than the relations by \citetalias{Taniguchi2018} as a result of our careful selection of the isolated \ion{Fe}{i} lines. 
We here apply the relations directly to ten nearby RSGs in the same way as to red giants in \autoref{ssec:ReDetTeffRGB}. 
\autoref{table:TLDRRSG} lists the resultant $T_{\mathrm{LDR}}$ of our targets together with the temperatures derived using the relations of \citetalias{Taniguchi2018}. 

The accuracy of our $T_{\mathrm{LDR}}$ of RSGs strongly depends on the assumption that empirical LDR--$T_{\mathrm{eff}}$ relations of \ion{Fe}{i} lines are well consistent between giants and supergiants. 
\citetalias{Jian2020} concluded that, at $T_{\mathrm{eff}}=4500$ and $5000\ur{K}$, the LDRs of \ion{Fe}{i} lines of dwarfs, giants and supergiants agree with each other. 
In order to test whether the agreement is also found between red giants and RSGs with $3500\lesssim T_{\mathrm{eff}}\lesssim 4000\ur{K}$, we first compare $T_{\mathrm{LDR}}$(TW) and $T_{\mathrm{LDR}}$(T18) to examine the dependence of the LDRs of multiple species on the surface gravity. 
The effect of the surface gravity on the LDRs is, at least to some extent, governed by the difference in the ionization energies, $\Delta \chi $, of the elements forming low- and high-excitation potential lines~\citepalias{Jian2020}. 
In fact, a weak anti-correlation of $\sim -80\ur{K/eV}$ between $\Delta \chi $ and $T_{\mathrm{eff}}$ from individual LDR relations ($T_{j}$) is seen for the ten RSGs~(\autoref{fig:IPDep}). 
This trend is similar to the one observed by \citetalias{Jian2020} for giants and supergiants at higher temperatures. 
It indicates that the surface gravity effect on the LDRs remains similar down to the low-temperature ranges of the RSGs and, therefore, $T_{\mathrm{LDR}}$(T18) of the RSGs have systematic errors. 
Moreover, the residuals from the $\Delta \chi $--$T_{j}$ correlation for most of the line pairs can be explained with the measurement error. 
Therefore, the combination of the $\Delta \chi $ effect and the measurement error may have a larger impact on $T_{j}$ estimation for most line pairs with multiple species than other effects (e.g. 3D non-LTE correction). 

\begin{figure}
\centering 
\includegraphics[width=\columnwidth ]{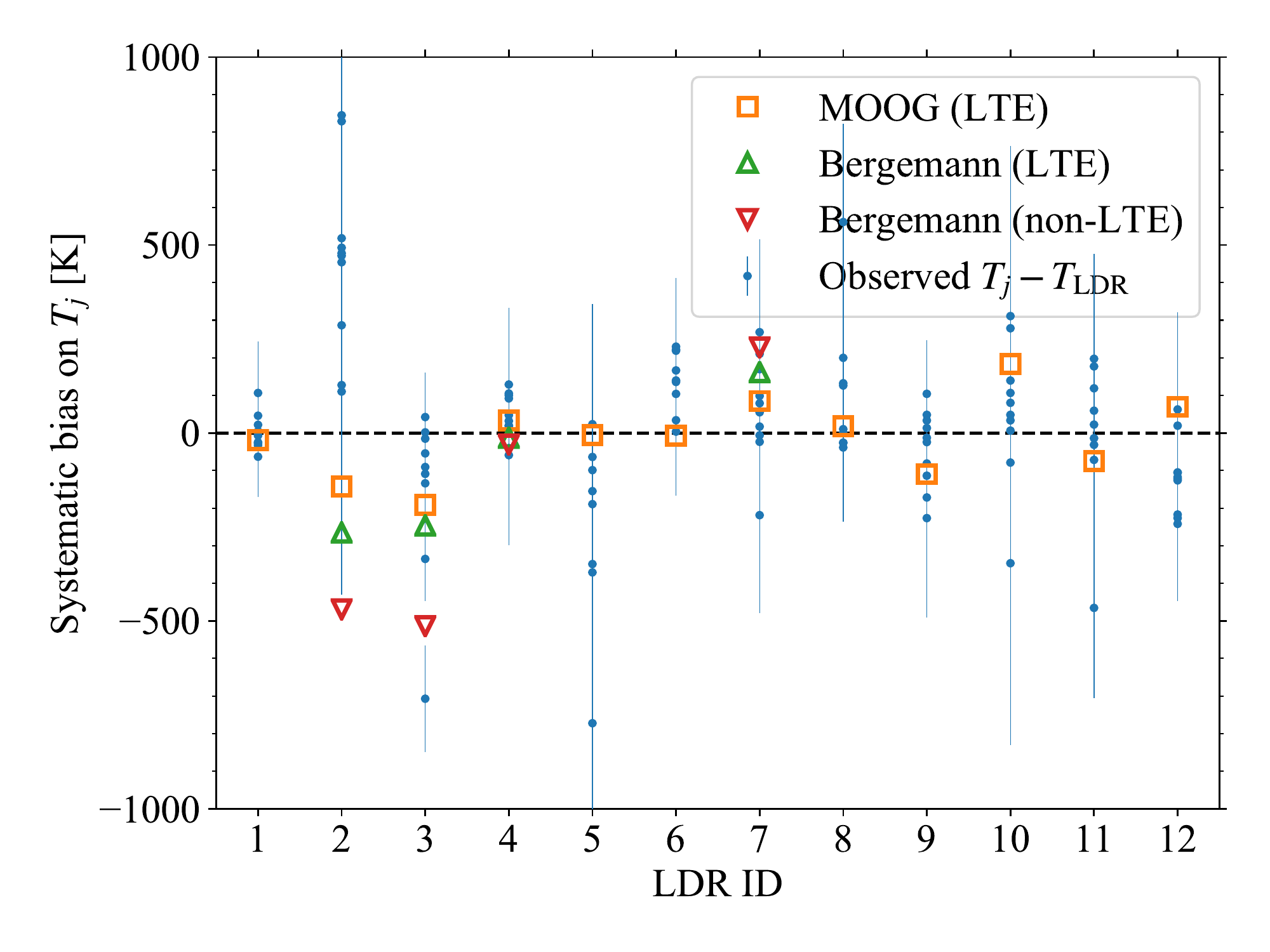}
\caption{Blue dots show the empirical deviation between the temperatures from a given line pair, $T_{j}$, and the LDR temperatures, $T_{\mathrm{LDR}}$(TW), for each \ion{Fe}{i} line pair for each star. Orange squares, green triangles and red inverted triangles show theoretically-estimated systematic bias~($a_{j}\Delta \log r_{j}$) of the LDR-based temperatures $T_{j}$ caused by the difference between red giants and RSGs calculated with MOOG, Bergemann's tool with LTE and that with non-LTE, respectively. }
\label{fig:CompDeltaTj}
\end{figure}

Concerning the error in $T_{\mathrm{LDR}}$, $\Delta T_{\mathrm{LDR}}^{\mathrm{Eq}\ref{eq:DeltaTLDR2}}$(T18) is found to be systematically~($1.5\text{--}2$ times) smaller than $\Delta T_{\mathrm{LDR}}^{\mathrm{Eq}\ref{eq:DeltaTLDR1}}$(T18); this fact suggests that $T_{j}$ for T18, hence $T_{\mathrm{LDR}}$(T18), has systematic errors introduced by the effect of the surface gravity. 
In contrast, $\Delta T_{\mathrm{LDR}}^{\mathrm{Eq}\ref{eq:DeltaTLDR2}}$(TW) is similar to or larger than $\Delta T_{\mathrm{LDR}}^{\mathrm{Eq}\ref{eq:DeltaTLDR1}}$(TW). 
This fact implies that the scatter of $T_{j}$ of \ion{Fe}{i} line pairs with each star can be explained mostly with the expected statistical errors except for a few cases; three stars~(TV Gem, BU Gem and \textpsi ${}^{1}$ Aur) have larger line broadening than the other RSGs, which could slightly bias their $T_{\mathrm{LDR}}$. 
However, the effect could be negligible because the lines that we used are well isolated ones. 
We also examine the error budget of $T_{\mathrm{LDR}}$(TW) using $T_{j}-T_{\mathrm{LDR}}$, which is a measure of the error in $T_{j}$~(blue dots in \autoref{fig:CompDeltaTj}). 
Though $T_{j}-T_{\mathrm{LDR}}$ with each \ion{Fe}{i} line pair has a distribution with a non-zero offset, the offsets are within $\sim \pm 150\ur{K}$ for the pairs except the pair ID~(2), which has an unpredictable large offset~($\sim 500\ur{K}$). 
These offsets are smaller than the measurement error, which indicates, again, that the error in $T_{\mathrm{LDR}}$(TW) is dominated by the expected statistical error. 
Moreover, the small systematic offsets may be partly explained by other lines contaminating the \ion{Fe}{i} lines that we use. 

Consequently, our new LDR--$T_{\mathrm{eff}}$ relations with only iron lines~($\Delta \chi =0\ur{eV}$) are not affected by the surface gravity. 
Since $\Delta T_{\mathrm{LDR}}^{\mathrm{Eq}\ref{eq:DeltaTLDR2}}$ may be slightly overestimated~(see the discussion on the `effective' S/N in \autoref{sssec:SNEstimateReal}), we adopt $\Delta T_{\mathrm{LDR}}^{\mathrm{Eq}\ref{eq:DeltaTLDR1}}$ as the more accurate error of $T_{\mathrm{LDR}}$ for the RSGs than $\Delta T_{\mathrm{LDR}}^{\mathrm{Eq}\ref{eq:DeltaTLDR2}}$. 
Note that since $\Delta T_{\mathrm{LDR}}^{\mathrm{Eq}\ref{eq:DeltaTLDR2}}$ is calculated using the scatter of $T_{j}$, this error includes a part of the systematic error on $T_{j}$. 
The inclusion of the systematic error and the larger measurement error may explain the larger $\Delta T_{\mathrm{LDR}}^{\mathrm{Eq}\ref{eq:DeltaTLDR2}}$(TW) of RSGs than that of cool red giants. 

In order to further examine the consistency between the LDR--$T_{\mathrm{eff}}$ relations of red giants and RSGs, we compare the \ion{Fe}{i} LDRs of the two groups by making use of their model spectra synthesized with MOOG~(see \autoref{sec:ConstructingLDR} for spectral synthesis). 
We calculated the differences $\Delta \log r_{j}$ in the logarithm of the LDRs $\log r_{j}$ between RSG3 and RGB1, which have the same $T_{\mathrm{eff}}$ of $3850\ur{K}$~(see \autoref{table:ImaginaryRSGAtmos} for their stellar parameters). 
Though $\Delta \log r_{j}$ calculated with the imperfect model spectra cannot reproduce observed one in a quantitative sense, $a_{j}\Delta \log r_{j}$ can be treated as a measure of the systematic error on $T_{j}$ due to the difference in the LDR--$T_{\mathrm{eff}}$ relations between red giants and RSGs. 
\autoref{fig:CompDeltaTj} shows $a_{j}\Delta \log r_{j}$ and $T_{j}-T_{\mathrm{LDR}}$ for each line pair. 
Except the pair ID~(2), $a_{j}\Delta \log r_{j}$ calculated with three types of synthetic spectra and observed $T_{j}-T_{\mathrm{LDR}}$ are similar to each other in a qualitative sense. 
Moreover, since $a_{j}\Delta \log r_{j}$ calculated with MOOG has a nearly symmetric distribution around the zero, the systematic errors for individual line pairs are expected to be cancelled out by taking an average of $T_{j}$. 
Though the reason of the inconsistency of the pair ID~(2) is unclear, it might be safe to reject this pair because this pair is suggested to involve a large~($\gtrsim 200\ur{K}$) systematic error both observationally and theoretically. 
We, thus, regard the weighted mean of $T_{j}$ after excluding the pair ID~(2) as the final estimate of the LDR temperature of RSGs~(\autoref{table:TLDRRSG_final}). 

Since some lines of red giants and RSGs suffer from the non-LTE effects in the extended atmospheres of these objects~\citep[e.g.][]{Bergemann2012b,Lind2012}, we finally examine how much the non-LTE effects could affect the LDR temperatures. 
For this purpose, we used the online spectral synthesis tool developed by M. Bergemann's group~\citep{Kovalev2019}\footnote{Last accessed on 2020 June 1. } to synthesize model spectra with and without the non-LTE effect. 
This tool can account for the non-LTE effect for many \ion{Fe}{i} lines calculated by \citet{Bergemann2012a}. 
However, the line list of the tool contains only half of the \ion{Fe}{i} lines that we have selected, and we could calculate the LDRs of only four of the final line pairs with the tool (IDs 2, 3, 4 and 7). 
The difference in $a_{j}\Delta \log r_{j}$ between LTE and non-LTE model spectra gives an estimate of the systematic bias of the LDR temperature caused by the non-LTE effect: $\sim 200\ur{K}$ for ID~(2), $\sim 250\ur{K}$ for ID~(3), $\sim 0\ur{K}$ for ID~(4) and $\sim 100\ur{K}$ for ID~(7)~(see \autoref{fig:CompDeltaTj}). 
While the observational results indicated by blue dots in \autoref{fig:CompDeltaTj} show consistent trends for the pairs ID~(4) and (7), we found conflicting results for the pairs ID~(2) and (3). 
The former two line pairs with the small non-LTE effect indicate that the systematic bias caused by the non-LTE effect is $\sim 100\ur{K}$ or less, though we cannot draw a strong conclusion based on the two pairs only. 
Moreover, it is hard to predict the systematic bias caused by the non-LTE effects without the non-LTE calculations for all the lines fully performed. 
We note that the 3D effect could also have an impact on line depths of late-type stars~\citep[e.g.][]{Collet2007,Kucinskas2013}, but the lack of a comprehensive grid of 3D model atmospheres of RSGs prevents us from estimating its impact. 

\begin{table*}
\centering 
\caption{Final estimates of effective temperatures and luminosities of ten RSGs derived with $11$ LDR--$T_{\mathrm{eff}}$ relations except ID~(2)~(see \autoref{ssec:TLDRRSG}). The statistical error associated with $T_{\mathrm{LDR}}$ is calculated with \autoref{eq:DeltaTLDR1}. Fourth and fifth columns show the parallax tabulated in the \textit{Gaia} EDR3 catalogue~\citep{Gaia2020} and the bias involved in it estimated following \citet{Lindegren2020b}, respectively, except Betelgeuse. Its parallax marked with an asterisk is from the \textsc{Hipparcos} catalogue~\citep{vanLeeuwen2007}. }
\label{table:TLDRRSG_final}
\begin{tabular}{lrlclcc}\toprule 
Object & HD & \multicolumn{1}{c}{$T_{\mathrm{LDR}} [K]$} & $N_{\mathrm{pair}}$ & \multicolumn{1}{c}{Parallax [mas]} & Bias [mas] & $\log (L/L_{\odot })$ \\ \midrule 
\textzeta \ Cep & 210745 & $4073\pm 31$ & $10$ & $3.297\pm 0.146$ & $-0.022$ & $3.73^{+0.08}_{-0.08}$ \\
41 Gem & 52005 & $3940\pm 29$ & $11$ & $0.721\pm 0.091$ & $-0.033$ & $4.28^{+0.17}_{-0.17}$ \\
\textxi \ Cyg & 200905 & $3891\pm 24$ & $11$ & $2.836\pm 0.127$ & $-0.023$ & $3.96^{+0.09}_{-0.09}$ \\
V809 Cas & 219978 & $3768\pm 35$ & $11$ & $0.999\pm 0.039$ & $-0.032$ & $4.57^{+0.08}_{-0.08}$ \\
V424 Lac & 216946 & $3749\pm 49$ & $11$ & $1.402\pm 0.113$ & $-0.027$ & $4.22^{+0.10}_{-0.10}$ \\
\textpsi ${}^{1}$ Aur & 44537 & $3740\pm 67$ & $11$ & $0.443\pm 0.110$ & $-0.036$ & $5.25^{+0.24}_{-0.20}$ \\
TV Gem & 42475 & $3676\pm 117$ & $8$ & $0.478\pm 0.135$ & $-0.029$ & $5.10^{+0.28}_{-0.22}$ \\
BU Gem & 42543 & $3883\pm 67$ & $9$ & $0.564\pm 0.125$ & $-0.043$ & $5.05^{+0.22}_{-0.19}$ \\
Betelgeuse & 39801 & $3611\pm 38$ & $11$ & $6.55\pm 0.83$${}^{\ast }$ &  & $4.91^{+0.14}_{-0.13}$ \\
NO Aur & 37536 & $3651\pm 31$ & $11$ & $0.919\pm 0.093$ & $-0.042$ & $4.49^{+0.12}_{-0.11}$ \\\bottomrule 
\end{tabular}
\end{table*}

\subsection{Comparison with previous $T_{\mathrm{eff}}$ determinations}\label{ssec:TLDRlit}

Betelgeuse is one of the most studied RSGs, whose $T_{\mathrm{eff}}$ is suggested to be free from strong time variability~\citep{White1978,Bester1996,Gray2008b,Levesque2020}, and thus it must be a good standard RSG for comparing $T_{\mathrm{eff}}$ by different methods. 
Many previous works have estimated $T_{\mathrm{eff}}$ of Betelgeuse using a variety of methods, e.g., broad-band interferometry~\citep[e.g.][]{Dyck1998,Haubois2009}, SED fitting~\citep[e.g.][]{Tsuji1976,Scargle1979}, the TiO method~\citep[e.g.][]{Levesque2005,Levesque2020} and the excitation balance of \ce{CO} molecular lines~\citep[e.g.][]{Lambert1984,Carr2000}, as summarized in \citet{Dolan2016}. 
These methods may, however, be affected by the systematic uncertainties described in \autoref{sec:IntroRSGTeff}. 
\citet{Davies2010} used the \textit{J}-band technique, which they claim is less affected than some other methods by these systematic uncertainties, and obtained $3520\pm 160$ and $3660\pm 170\ur{K}$ for spectra with spectral resolutions of $R=2,000$ and $6,000$, respectively. 
Our result of $3611\pm 38\ur{K}$ for Betelgeuse is consistent with theirs. 
Their estimate for V424~Lac, $3580\pm 230\ur{K}$, based on its spectrum with $R=2,000$ is also consistent with ours, $3749\pm 49\ur{K}$, within the errors, but the variability of V424~Lac is poorly known. 
We note that $T_{\mathrm{eff}}$ of no other RSGs in our sample were measured by \citet{Davies2010} and other recent works using the \textit{J}-band technique. 

Another promising approach is the spectro-interferometric observation, which is less affected by the outer envelopes of RSGs. 
\citet{Ohnaka2011} determined Betelgeuse's $T_{\mathrm{eff}}$ to be $3690\pm 54\ur{K}$ with this approach. 
In addition \citet{ArroyoTorres2013} estimated it to be $3620\pm 137\ur{K}$, using the angular diameter measured by \citet{Ohnaka2011}. 
Our result of $3611\pm 38\ur{K}$ for Betelgeuse is consistent with theirs, and hence supports the validity of the $T_{\mathrm{eff}}$ scale that we obtained. 
Unfortunately, no other RSGs than Betelgeuse in our sample have the spectro-interferometric $T_{\mathrm{eff}}$ in literature to compare with. 

\begin{figure}
\centering 
\includegraphics[width=\columnwidth ]{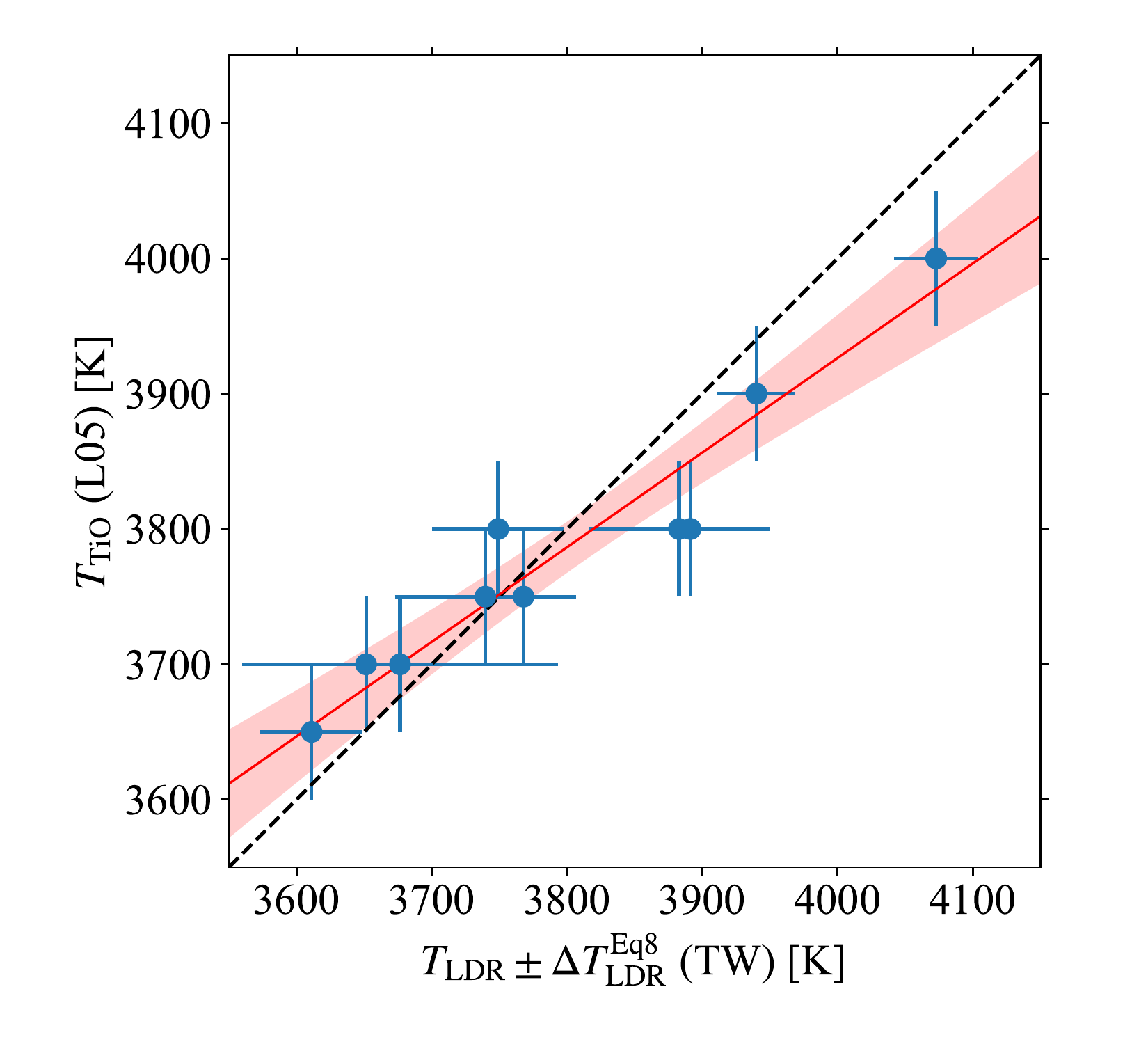}
\caption{Comparison between the effective temperatures of RSGs determined in this work~(horizontal axis) and those by \citet{Levesque2005}~(vertical axis). A red line shows the linear regression line of blue points with the light red shade showing the $1\sigma $ confidence intervals. }
\label{fig:CompTLDRLevesque}
\end{figure}

\autoref{fig:CompTLDRLevesque} compares the effective temperatures of all the RSGs in this work~($T_{\mathrm{LDR}}\pm \Delta T_{\mathrm{LDR}}^{\mathrm{Eq}\ref{eq:DeltaTLDR1}}$) and those determined by \citet{Levesque2005}~($T_{\mathrm{TiO}}$(L05)), where the \ce{TiO} method was employed. 
Whereas comparing $T_{\mathrm{eff}}$ of each star is not so practical, given that the time variations of $T_{\mathrm{eff}}$ for RSGs are as large as several hundreds kelvins in the worst cases~\citep[e.g.][]{Levesque2007,Clark2010,Wasatonic2015}, the comparison of these as a whole tells something significant; the two sets of $T_{\mathrm{eff}}$ were found to be well consistent, considering the errors of the former~($\Delta T_{\mathrm{LDR}}^{\mathrm{Eq}\ref{eq:DeltaTLDR1}}$) and latter~($\sim 50\ur{K}$), but with a slope of $0.70\pm 0.14$, which is slightly different from one. 
This consistency indicates that the \ce{TiO} method yields not strongly biased $T_{\mathrm{eff}}$ of RSGs that have the solar abundance pattern. 
However, it is unclear whether the \ce{TiO} method can yield unbiased $T_{\mathrm{eff}}$ of RSGs that have non-solar abundances, considering the uncertainties discussed in \autoref{sec:IntroRSGTeff}. 
In fact, \citet{Levesque2006} and \citet{Davies2013} determined $T_{\mathrm{eff}}$ of common $\sim 20$ RSGs in the Magellanic Clouds, using the \ce{TiO} method, but $T_{\mathrm{eff}}$ by \citet{Levesque2006} are systematically $\sim 100\ur{K}$ higher than those by \citet{Davies2013}. 
Moreover, \citet{Davies2013} compared $T_{\mathrm{eff}}$ derived with the \ce{TiO} method to that with the SED method, and found that the latter method yielded $\sim 400\ur{K}$ higher $T_{\mathrm{eff}}$ than the former.

\subsection{Red supergiants on the HR diagram}\label{ssec:HRD}

In many previous works, more than a decade ago, observationally-determined $T_{\mathrm{eff}}$ of the brightest RSGs~($M_{\mathrm{bol}}<-7\ur{mag}$) were often lower than expected from theoretical models at a given luminosity~\citep[see, e.g., Figure 8 in][]{Massey2003a}. 
Since then, several observational studies have compared their $T_{\mathrm{eff}}$ with theoretical models, mainly with Geneva's stellar evolution model~\citep{Ekstrom2012,Georgy2013}, on the Hertzsprung-Russel~(HR) diagram. Some of them found good consistency between them~\citep[e.g.][]{Levesque2005,Davies2013,Wittkowski2017}, whereas some others found offsets between the measurements and theoretical models~\citep[e.g.][]{Levesque2006,Tabernero2018}. 

In order to plot our observed RSGs on the HR diagram, we derived the bolometric luminosity, using the parallax and bolometric magnitude of each source as follows. 
As for the parallaxes, we used the values in \textit{Gaia} EDR3~\citep{Gaia2016,Gaia2020} for all our RSG samples except Betelgeuse, for which we used the parallax in the \textsc{Hipparcos} catalogue by \citet{vanLeeuwen2007} because \textit{Gaia} EDR3 has no entry for it. 
Since there is a systematic bias in the \textit{Gaia} parallax~\citep{Lindegren2020a}, we followed the recipe by \citet{Lindegren2020b} to reduce the bias. 
In brief, we inputted $G$-band magnitude, effective wavenumber and ecliptic latitude tabulated in \textit{Gaia} EDR3 into the code developed by them\footnote{\url{https://gitlab.com/icc-ub/public/gaiadr3_zeropoint}} to calculate the bias estimate. 
Estimated biases, summarized in \autoref{table:TLDRRSG_final} and ranging from $-43$ to $-22\,\text{\textmu as}$, only change the final $\log L$ by up to $0.07$, which is smaller than the statistical error in it. 
Nevertheless, it should be kept in mind that (i)~all the sample RSGs have $G<6\ur{mag}$, which is out of the range of the bias calibration by \citet{Lindegren2020b}, and (ii)~the \textit{Gaia} parallaxes of very bright stars could contain large systematic biases; for example, several stars with $G\lesssim 4\ur{mag}$ and with parallaxes larger than $5\ur{mas}$ have systematic biases of $\gtrsim 1\ur{mas}$ in \textit{Gaia} DR2~\citep{Drimmel2019} though the bias would be smaller in EDR3~\citep{Torra2020}, and \textit{Gaia} DR3 parallaxes of all our sample RSGs are smaller than $4\ur{mas}$. 
We note that the statistical errors of the parallaxes of our targets tend to be larger than those expected from their magnitudes in both the \textsc{Hipparcos}~\citep{vanLeeuwen2007} and \textit{Gaia} EDR3~\citep{Lindegren2020a} catalogues. 
It is most likely, at least partly, due to strong granulation in RSGs, which fluctuates the positions of their photometric centroids by $\sim 0.1\ur{AU}$~\citep{Chiavassa2011a,Pasquato2011}. 

As for the bolometric magnitude, we estimated it for each RSG sample from the 2MASS \textit{K}$_{\mathrm{s}}$-band magnitude~\citep{Skrutskie2006}, considering the extinction and the bolometric correction.
The sampled epochs of the catalogued \textit{K}$_{\mathrm{s}}$-band magnitude that we used are different from those of their WINERED observations. 
However, the difference in the epochs would not be a significant problem, given that the magnitude variation of the RSG in the infrared bands is known to be negligible~\citep[e.g.][]{Yang2018,Ren2019}, in contrast to the optical variation, which is as large as a few magnitudes~\citep[and references therein]{Kiss2006,Soraisam2018}. 
Concerning the interstellar and circumstellar extinction, we converted the \textit{V}-band extinction $A(V)$ of the RSGs measured by \citet{Levesque2005}, with the typical precision of $0.15\ur{mag}$, to the \textit{K}$_{\mathrm{s}}$-band extinction $A(K_{\mathrm{s}})$, using the reddening law $A(K_{\mathrm{s}})/A(V)=0.116$ by \citet{Cardelli1989}, assuming the total-to-selective extinction ratio $R_{V}=3.1$ as in \citet{Levesque2005}. 
We estimated the bolometric correction for the \textit{K}$_{\mathrm{s}}$ band, $\mathrm{BC}_{K_{\mathrm{s}}}$, interpolating the $\mathrm{BC}_{K}$ value with regard to the obtained $T_{\mathrm{eff}}$~(\autoref{ssec:TLDRRSG}) in the tabulated relation between $T_{\mathrm{eff}}$ and $\mathrm{BC}_{K}$ for RSGs in Table~6 of \citet{Levesque2005}. 
Since the extinction to our sample RSGs are small, $A(V)=2.17\ur{mag}$ at most, the choice of the reddening law and the $R_{V}$ value and the difference between \textit{K}$_{\mathrm{s}}$ and \textit{K} do not significantly affect the resultant luminosities. 
Finally, the bolometric luminosity~($L$) of each target RSG and its error were calculated using the Monte Carlo method~(\autoref{table:TLDRRSG_final}). 

\begin{figure}
\centering 
\includegraphics[width=\columnwidth ]{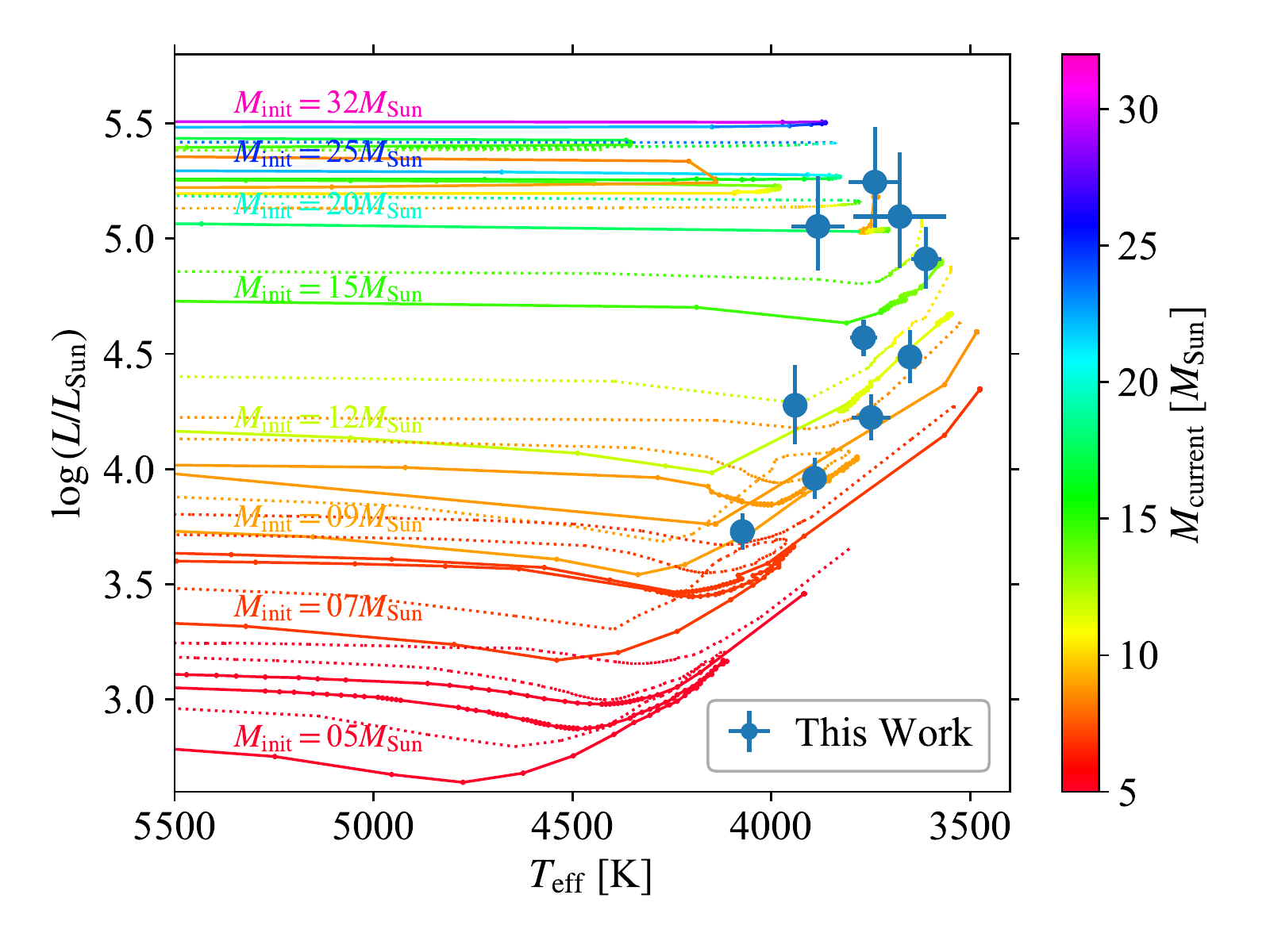}
\caption{HR diagram of RSGs. Blue circles show the effective temperatures and bolometric luminosities of the sample RSGs determined in this work. The solar-metallicity Geneva evolutionary track by \citet{Ekstrom2012} with and without rotation are shown with dotted and solid lines, respectively, colour-coded with the current mass. }
\label{fig:HRdiagramRSG}
\end{figure}

In the HR diagram in \autoref{fig:HRdiagramRSG}, we compared the distribution of the data points of our sample RSGs in our estimates with those expected from the latest Geneva's stellar evolution model with the solar metallicity~\citep{Ekstrom2012}. 
Although the sample size is limited, $T_{\mathrm{eff}}$ obtained in this work are consistent with the range of $T_{\mathrm{eff}}$ in which RSGs are expected to stay relatively long time. 
A larger sample of RSGs, especially those with a higher luminosity, would enable us to investigate the properties of the Galactic RSGs, e.g., comparing the $T_{\mathrm{eff}}$ distribution more closely with various evolutionary models, determining the relation between the spectral type and $T_{\mathrm{eff}}$ and so on.

\section{Summary and Future Prospects}

In this paper, we calibrated the empirical relations between twelve LDRs of \ion{Fe}{i} lines and $T_{\mathrm{eff}}$, using nine solar-metal red giants observed with WINERED. 
These relations enabled us to determine $T_{\mathrm{eff}}$ of red giants to a precision of $\sim 30\ur{K}$ in the best cases, i.e., early-M type giants with good S/N. 
We applied these relations to ten nearby RSGs and obtained $T_{\mathrm{eff}}$ to a precision of $\sim 40\ur{K}$. 
The estimated $T_{\mathrm{eff}}$ are in good agreement with the values estimated with different methods in relevant literature: those with the \ce{TiO} method by \citet{Levesque2005} and that of Betelgeuse on the basis of spectro-interferometry by \citet{Ohnaka2011} and \citet{ArroyoTorres2013}, as well as those expected in Geneva's stellar evolution model~\citep{Ekstrom2012}. 

Our method uses only \ion{Fe}{i} lines in the near-infrared~(the \textit{Y} and \textit{J} bands) high-resolution spectra. Because of it, our method is expected to give more unbiased $T_{\mathrm{eff}}$ of RSGs than the other published spectroscopic methods, which rely on lines of molecules and/or multiple atoms and inevitably suffer from some significant systematic uncertainties. 
First, our method is independent of chemical abundance ratios because our method relies on only one species, \ion{Fe}{i}. 
Second, our method is expected to be less affected by the granulation than the methods using molecular bands because the granulation tends to vary the temperature structure of the upper atmospheric layers in particular and, hence, the strengths of molecular bands. 
Finally, our method is less affected by the surface gravity effect from which conventional LDR methods with multiple species suffer. 

Though there are possible systematic biases of the temperatures derived with individual line pairs due to the surface gravity, microturbulent velocity and/or line broadening effects, it is expected that they cancel each other out when the temperatures derived with several pairs are averaged. 
In contrast, it is unclear whether the systematic bias due to the non-LTE effect, as large as $\sim 250\ur{K}$ for each pair, can be reduced by taking the average of individual estimates. 
Nevertheless, the final LDR temperatures using all the $11$ pairs are well consistent within $\sim 100\ur{K}$ with the temperatures derived with the LDR ID (4), which is insensitive to the non-LTE effect. 
This fact indicates that the actual systematic bias due to the non-LTE effect is as small as $\sim 100\ur{K}$. 
To further examine the possible systematic errors, in-depth consideration with reliable 3D non-LTE model spectra is desired. 

Our sample is limited to solar-metal objects, red giants and RSGs, and the relations we obtained in this work can be applied to solar-metal objects only. 
In order to get rid of this limitation, the metallicity dependence of the LDRs in the \textit{Y} and \textit{J} bands needs to be examined. 
Observations with recent and/or future near-infrared high-resolution spectrographs with high sensitivities~\citep[e.g. WINERED;][]{Ikeda2016,Ikeda2018} will provide sufficient-statistics data of RSGs at large distances, e.g., those in the inner/outer-Galaxy and some local-group galaxies like the Magellanic Clouds and more. 
Then $T_{\mathrm{eff}}$ of these sources can be determined in this method, whereas interferometric measurements of such distant sources would be practically impossible. 
Obtaining $T_{\mathrm{eff}}$ of RSGs in various environments is desirable to examine the potential environmental dependency of $T_{\mathrm{eff}}$ on [Fe/H] and to test stellar evolution models further.

\section*{Acknowledgements}
We acknowledge useful comments from the referee, Maria Bergemann. 
We are grateful to Masaomi Kinoshita, Yuki Moritani, Kenshi Nakanishi, Tetsuya Nakaoka, Kyoko Sakamoto and Yoshiharu Shinnaka for observing a part of our targets. 
We also thank the staff of Koyama Astronomical Observatory for their support during our observations. 
The WINERED was developed by the University of Tokyo and the Laboratory of Infrared High-resolution spectroscopy~(LiH), Kyoto Sangyo University under the financial supports of Grants-in-Aid, KAKENHI, from Japan Society for the Promotion of Science~(JSPS; Nos. 16684001, 20340042, and 21840052) and the MEXT Supported Program for the Strategic Research Foundation at Private Universities~(Nos. S0801061 and S1411028). 
This work has been supported by Masason Foundation. 
DT acknowledges financial support from Toyota/Dwango AI scholarship and Iwadare Scholarship Foundation in 2020. 
NM, NK and HK acknowledge financial support of KAKENHI No. 18H01248. 
NK also acknowledges support through the Japan--India Science Cooperative Program between 2013 and 2018 under agreement between the JSPS and the Department of Science and Technology~(DST) in India. 
KF acknowledges financial support of KAKENHI No. 16H07323. 
HS acknowledges financial support of KAKENHI No. 19K03917. 

This work has made use of the VALD database, operated at Uppsala University, the Institute of Astronomy RAS in Moscow, and the University of Vienna. 
This research has made use of the SIMBAD database, operated at CDS, Strasbourg, France. 
This work has made use of data from the European Space Agency~(ESA) mission \textit{Gaia}~(\url{https://www.cosmos.esa.int/gaia}), processed by the \textit{Gaia} Data Processing and Analysis Consortium~(DPAC, \url{https://www.cosmos.esa.int/web/gaia/dpac/consortium}). Funding for the DPAC has been provided by national institutions, in particular the institutions participating in the \textit{Gaia} Multilateral Agreement. 
This publication makes use of data products from the Two Micron All Sky Survey, which is a joint project of the University of Massachusetts and the Infrared Processing and Analysis Center/California Institute of Technology, funded by the National Aeronautics and Space Administration and the National Science Foundation.

\section*{Data Availability}
The data underlying this article will be shared on reasonable request to the corresponding author.

\bibliographystyle{mnras}
\bibliography{LDRFe_taniguchi}

\begin{thebibliography}{}
\makeatletter
\relax
\def\mn@urlcharsother{\let\do\@makeother \do\$\do\&\do\#\do\^\do\_\do\%\do\~}
\def\mn@doi{\begingroup\mn@urlcharsother \@ifnextchar [ {\mn@doi@}
  {\mn@doi@[]}}
\def\mn@doi@[#1]#2{\def\@tempa{#1}\ifx\@tempa\@empty \href
  {http://dx.doi.org/#2} {doi:#2}\else \href {http://dx.doi.org/#2} {#1}\fi
  \endgroup}
\def\mn@eprint#1#2{\mn@eprint@#1:#2::\@nil}
\def\mn@eprint@arXiv#1{\href {http://arxiv.org/abs/#1} {{\tt arXiv:#1}}}
\def\mn@eprint@dblp#1{\href {http://dblp.uni-trier.de/rec/bibtex/#1.xml}
  {dblp:#1}}
\def\mn@eprint@#1:#2:#3:#4\@nil{\def\@tempa {#1}\def\@tempb {#2}\def\@tempc
  {#3}\ifx \@tempc \@empty \let \@tempc \@tempb \let \@tempb \@tempa \fi \ifx
  \@tempb \@empty \def\@tempb {arXiv}\fi \@ifundefined
  {mn@eprint@\@tempb}{\@tempb:\@tempc}{\expandafter \expandafter \csname
  mn@eprint@\@tempb\endcsname \expandafter{\@tempc}}}

\bibitem[\protect\citeauthoryear{{Alonso-Santiago}, {Negueruela}, {Marco},
  {Tabernero}, {Gonz{\'a}lez-Fern{\'a}ndez}  \& {Castro}}{{Alonso-Santiago}
  et~al.}{2019}]{AlonsoSantiago2019}
{Alonso-Santiago} J.,  {Negueruela} I.,  {Marco} A.,  {Tabernero} H.~M.,
  {Gonz{\'a}lez-Fern{\'a}ndez} C.,   {Castro} N.,  2019, \mn@doi [\aap]
  {10.1051/0004-6361/201936109}, \href
  {https://ui.adsabs.harvard.edu/abs/2019A&A...631A.124A} {631, A124}

\bibitem[\protect\citeauthoryear{{Arroyo-Torres}, {Wittkowski}, {Marcaide}  \&
  {Hauschildt}}{{Arroyo-Torres} et~al.}{2013}]{ArroyoTorres2013}
{Arroyo-Torres} B.,  {Wittkowski} M.,  {Marcaide} J.~M.,   {Hauschildt} P.~H.,
  2013, \mn@doi [\aap] {10.1051/0004-6361/201220920}, \href
  {https://ui.adsabs.harvard.edu/abs/2013A&A...554A..76A} {554, A76}

\bibitem[\protect\citeauthoryear{{Asplund}, {Grevesse}, {Sauval}  \&
  {Scott}}{{Asplund} et~al.}{2009}]{Asplund2009}
{Asplund} M.,  {Grevesse} N.,  {Sauval} A.~J.,   {Scott} P.,  2009, \mn@doi
  [\araa] {10.1146/annurev.astro.46.060407.145222}, \href
  {https://ui.adsabs.harvard.edu/abs/2009ARA&A..47..481A} {47, 481}

\bibitem[\protect\citeauthoryear{{Bergemann}, {Lind}, {Collet}, {Magic}  \&
  {Asplund}}{{Bergemann} et~al.}{2012a}]{Bergemann2012a}
{Bergemann} M.,  {Lind} K.,  {Collet} R.,  {Magic} Z.,   {Asplund} M.,  2012a,
  \mn@doi [\mnras] {10.1111/j.1365-2966.2012.21687.x}, \href
  {https://ui.adsabs.harvard.edu/abs/2012MNRAS.427...27B} {427, 27}

\bibitem[\protect\citeauthoryear{{Bergemann}, {Kudritzki}, {Plez}, {Davies},
  {Lind}  \& {Gazak}}{{Bergemann} et~al.}{2012b}]{Bergemann2012b}
{Bergemann} M.,  {Kudritzki} R.-P.,  {Plez} B.,  {Davies} B.,  {Lind} K.,
  {Gazak} Z.,  2012b, \mn@doi [\apj] {10.1088/0004-637X/751/2/156}, \href
  {https://ui.adsabs.harvard.edu/abs/2012ApJ...751..156B} {751, 156}

\bibitem[\protect\citeauthoryear{{Bester}, {Danchi}, {Hale}, {Townes},
  {Degiacomi}, {Mekarnia}  \& {Geballe}}{{Bester} et~al.}{1996}]{Bester1996}
{Bester} M.,  {Danchi} W.~C.,  {Hale} D.,  {Townes} C.~H.,  {Degiacomi} C.~G.,
  {Mekarnia} D.,   {Geballe} T.~R.,  1996, \mn@doi [\apj] {10.1086/177246},
  \href {https://ui.adsabs.harvard.edu/abs/1996ApJ...463..336B} {463, 336}

\bibitem[\protect\citeauthoryear{{Biazzo}, {Frasca}, {Catalano}  \&
  {Marilli}}{{Biazzo} et~al.}{2007}]{Biazzo2007}
{Biazzo} K.,  {Frasca} A.,  {Catalano} S.,   {Marilli} E.,  2007, \mn@doi
  [Astronomische Nachrichten] {10.1002/asna.200710781}, \href
  {https://ui.adsabs.harvard.edu/abs/2007AN....328..938B} {328, 938}

\bibitem[\protect\citeauthoryear{{Blanco-Cuaresma}, {Soubiran}, {Jofr{\'e}}  \&
  {Heiter}}{{Blanco-Cuaresma} et~al.}{2014}]{BlancoCuaresma2014}
{Blanco-Cuaresma} S.,  {Soubiran} C.,  {Jofr{\'e}} P.,   {Heiter} U.,  2014,
  \mn@doi [\aap] {10.1051/0004-6361/201323153}, \href
  {https://ui.adsabs.harvard.edu/abs/2014A&A...566A..98B} {566, A98}

\bibitem[\protect\citeauthoryear{{Brooke}, {Ram}, {Western}, {Li}, {Schwenke}
  \& {Bernath}}{{Brooke} et~al.}{2014}]{Brooke2014}
{Brooke} J. S.~A.,  {Ram} R.~S.,  {Western} C.~M.,  {Li} G.,  {Schwenke} D.~W.,
    {Bernath} P.~F.,  2014, \mn@doi [\apjs] {10.1088/0067-0049/210/2/23}, \href
  {https://ui.adsabs.harvard.edu/abs/2014ApJS..210...23B} {210, 23}

\bibitem[\protect\citeauthoryear{{Cardelli}, {Clayton}  \& {Mathis}}{{Cardelli}
  et~al.}{1989}]{Cardelli1989}
{Cardelli} J.~A.,  {Clayton} G.~C.,   {Mathis} J.~S.,  1989, \mn@doi [\apj]
  {10.1086/167900}, \href
  {https://ui.adsabs.harvard.edu/abs/1989ApJ...345..245C} {345, 245}

\bibitem[\protect\citeauthoryear{{Carr}, {Sellgren}  \& {Balachandran}}{{Carr}
  et~al.}{2000}]{Carr2000}
{Carr} J.~S.,  {Sellgren} K.,   {Balachandran} S.~C.,  2000, \mn@doi [\apj]
  {10.1086/308340}, \href
  {https://ui.adsabs.harvard.edu/abs/2000ApJ...530..307C} {530, 307}

\bibitem[\protect\citeauthoryear{{Chiavassa}, {Plez}, {Josselin}  \&
  {Freytag}}{{Chiavassa} et~al.}{2009}]{Chiavassa2009}
{Chiavassa} A.,  {Plez} B.,  {Josselin} E.,   {Freytag} B.,  2009, \mn@doi
  [\aap] {10.1051/0004-6361/200911780}, \href
  {https://ui.adsabs.harvard.edu/abs/2009A&A...506.1351C} {506, 1351}

\bibitem[\protect\citeauthoryear{{Chiavassa} et~al.,}{{Chiavassa}
  et~al.}{2011a}]{Chiavassa2011a}
{Chiavassa} A.,  et~al., 2011a, \mn@doi [\aap] {10.1051/0004-6361/201015768},
  \href {https://ui.adsabs.harvard.edu/abs/2011A&A...528A.120C} {528, A120}

\bibitem[\protect\citeauthoryear{{Chiavassa}, {Freytag}, {Masseron}  \&
  {Plez}}{{Chiavassa} et~al.}{2011b}]{Chiavassa2011b}
{Chiavassa} A.,  {Freytag} B.,  {Masseron} T.,   {Plez} B.,  2011b, \mn@doi
  [\aap] {10.1051/0004-6361/201117463}, \href
  {https://ui.adsabs.harvard.edu/abs/2011A&A...535A..22C} {535, A22}

\bibitem[\protect\citeauthoryear{{Choi}, {Dotter}, {Conroy}, {Cantiello},
  {Paxton}  \& {Johnson}}{{Choi} et~al.}{2016}]{Choi2016}
{Choi} J.,  {Dotter} A.,  {Conroy} C.,  {Cantiello} M.,  {Paxton} B.,
  {Johnson} B.~D.,  2016, \mn@doi [\apj] {10.3847/0004-637X/823/2/102}, \href
  {https://ui.adsabs.harvard.edu/abs/2016ApJ...823..102C} {823, 102}

\bibitem[\protect\citeauthoryear{{Clark}, {Ritchie}  \& {Negueruela}}{{Clark}
  et~al.}{2010}]{Clark2010}
{Clark} J.~S.,  {Ritchie} B.~W.,   {Negueruela} I.,  2010, \mn@doi [\aap]
  {10.1051/0004-6361/200913820}, \href
  {https://ui.adsabs.harvard.edu/abs/2010A&A...514A..87C} {514, A87}

\bibitem[\protect\citeauthoryear{{Coelho}, {Barbuy}, {Mel{\'e}ndez}, {Schiavon}
   \& {Castilho}}{{Coelho} et~al.}{2005}]{Coelho2005}
{Coelho} P.,  {Barbuy} B.,  {Mel{\'e}ndez} J.,  {Schiavon} R.~P.,   {Castilho}
  B.~V.,  2005, \mn@doi [\aap] {10.1051/0004-6361:20053511}, \href
  {https://ui.adsabs.harvard.edu/abs/2005A&A...443..735C} {443, 735}

\bibitem[\protect\citeauthoryear{{Collet}, {Asplund}  \& {Trampedach}}{{Collet}
  et~al.}{2007}]{Collet2007}
{Collet} R.,  {Asplund} M.,   {Trampedach} R.,  2007, \mn@doi [\aap]
  {10.1051/0004-6361:20066321}, \href
  {https://ui.adsabs.harvard.edu/abs/2007A&A...469..687C} {469, 687}

\bibitem[\protect\citeauthoryear{{Cunha}, {Sellgren}, {Smith}, {Ramirez},
  {Blum}  \& {Terndrup}}{{Cunha} et~al.}{2007}]{Cunha2007}
{Cunha} K.,  {Sellgren} K.,  {Smith} V.~V.,  {Ramirez} S.~V.,  {Blum} R.~D.,
  {Terndrup} D.~M.,  2007, \mn@doi [\apj] {10.1086/521813}, \href
  {https://ui.adsabs.harvard.edu/abs/2007ApJ...669.1011C} {669, 1011}

\bibitem[\protect\citeauthoryear{{Davies}, {Kudritzki}  \& {Figer}}{{Davies}
  et~al.}{2010}]{Davies2010}
{Davies} B.,  {Kudritzki} R.-P.,   {Figer} D.~F.,  2010, \mn@doi [\mnras]
  {10.1111/j.1365-2966.2010.16965.x}, \href
  {https://ui.adsabs.harvard.edu/abs/2010MNRAS.407.1203D} {407, 1203}

\bibitem[\protect\citeauthoryear{{Davies} et~al.,}{{Davies}
  et~al.}{2013}]{Davies2013}
{Davies} B.,  et~al., 2013, \mn@doi [\apj] {10.1088/0004-637X/767/1/3}, \href
  {https://ui.adsabs.harvard.edu/abs/2013ApJ...767....3D} {767, 3}

\bibitem[\protect\citeauthoryear{{Davies}, {Kudritzki}, {Gazak}, {Plez},
  {Bergemann}, {Evans}  \& {Patrick}}{{Davies} et~al.}{2015}]{Davies2015}
{Davies} B.,  {Kudritzki} R.-P.,  {Gazak} Z.,  {Plez} B.,  {Bergemann} M.,
  {Evans} C.,   {Patrick} L.,  2015, \mn@doi [\apj]
  {10.1088/0004-637X/806/1/21}, \href
  {https://ui.adsabs.harvard.edu/abs/2015ApJ...806...21D} {806, 21}

\bibitem[\protect\citeauthoryear{{Dolan}, {Mathews}, {Lam}, {Quynh Lan},
  {Herczeg}  \& {Dearborn}}{{Dolan} et~al.}{2016}]{Dolan2016}
{Dolan} M.~M.,  {Mathews} G.~J.,  {Lam} D.~D.,  {Quynh Lan} N.,  {Herczeg}
  G.~J.,   {Dearborn} D. S.~P.,  2016, \mn@doi [\apj]
  {10.3847/0004-637X/819/1/7}, \href
  {https://ui.adsabs.harvard.edu/abs/2016ApJ...819....7D} {819, 7}

\bibitem[\protect\citeauthoryear{{Drimmel}, {Bucciarelli}  \& {Inno}}{{Drimmel}
  et~al.}{2019}]{Drimmel2019}
{Drimmel} R.,  {Bucciarelli} B.,   {Inno} L.,  2019, \mn@doi [Research Notes of
  the American Astronomical Society] {10.3847/2515-5172/ab2632}, \href
  {https://ui.adsabs.harvard.edu/abs/2019RNAAS...3...79D} {3, 79}

\bibitem[\protect\citeauthoryear{{Drout}, {Massey}  \& {Meynet}}{{Drout}
  et~al.}{2012}]{Drout2012}
{Drout} M.~R.,  {Massey} P.,   {Meynet} G.,  2012, \mn@doi [\apj]
  {10.1088/0004-637X/750/2/97}, \href
  {https://ui.adsabs.harvard.edu/abs/2012ApJ...750...97D} {750, 97}

\bibitem[\protect\citeauthoryear{{Dyck} \& {Nordgren}}{{Dyck} \&
  {Nordgren}}{2002}]{Dyck2002}
{Dyck} H.~M.,  {Nordgren} T.~E.,  2002, \mn@doi [\aj] {10.1086/341039}, \href
  {https://ui.adsabs.harvard.edu/abs/2002AJ....124..541D} {124, 541}

\bibitem[\protect\citeauthoryear{{Dyck}, {van Belle}  \& {Thompson}}{{Dyck}
  et~al.}{1998}]{Dyck1998}
{Dyck} H.~M.,  {van Belle} G.~T.,   {Thompson} R.~R.,  1998, \mn@doi [\aj]
  {10.1086/300453}, \href
  {https://ui.adsabs.harvard.edu/abs/1998AJ....116..981D} {116, 981}

\bibitem[\protect\citeauthoryear{{Ekstr{\"o}m} et~al.,}{{Ekstr{\"o}m}
  et~al.}{2012}]{Ekstrom2012}
{Ekstr{\"o}m} S.,  et~al., 2012, \mn@doi [\aap] {10.1051/0004-6361/201117751},
  \href {https://ui.adsabs.harvard.edu/abs/2012A&A...537A.146E} {537, A146}

\bibitem[\protect\citeauthoryear{{Fraser} et~al.,}{{Fraser}
  et~al.}{2011}]{Fraser2011}
{Fraser} M.,  et~al., 2011, \mn@doi [\mnras]
  {10.1111/j.1365-2966.2011.19370.x}, \href
  {https://ui.adsabs.harvard.edu/abs/2011MNRAS.417.1417F} {417, 1417}

\bibitem[\protect\citeauthoryear{{Fukue} et~al.,}{{Fukue}
  et~al.}{2015}]{Fukue2015}
{Fukue} K.,  et~al., 2015, \mn@doi [\apj] {10.1088/0004-637X/812/1/64}, \href
  {https://ui.adsabs.harvard.edu/abs/2015ApJ...812...64F} {812, 64}

\bibitem[\protect\citeauthoryear{{Gaia Collaboration} et~al.,}{{Gaia
  Collaboration} et~al.}{2016}]{Gaia2016}
{Gaia Collaboration} et~al., 2016, \mn@doi [\aap]
  {10.1051/0004-6361/201629272}, \href
  {https://ui.adsabs.harvard.edu/abs/2016A&A...595A...1G} {595, A1}

\bibitem[\protect\citeauthoryear{{Gaia Collaboration} et~al.,}{{Gaia
  Collaboration} et~al.}{2021}]{Gaia2020}
{Gaia Collaboration} et~al., 2021, \mn@doi [\aap]
  {10.1051/0004-6361/202039657}, \href
  {https://ui.adsabs.harvard.edu/abs/2021A&A...649A...1G} {649, A1}

\bibitem[\protect\citeauthoryear{{Georgy} et~al.,}{{Georgy}
  et~al.}{2013}]{Georgy2013}
{Georgy} C.,  et~al., 2013, \mn@doi [\aap] {10.1051/0004-6361/201322178}, \href
  {https://ui.adsabs.harvard.edu/abs/2013A&A...558A.103G} {558, A103}

\bibitem[\protect\citeauthoryear{{Gray}}{{Gray}}{2008a}]{Gray2008a}
{Gray} D.~F.,  2008a, {The Observation and Analysis of Stellar Photospheres}.
Cambridge Univ. Press, Cambridge

\bibitem[\protect\citeauthoryear{{Gray}}{{Gray}}{2008b}]{Gray2008b}
{Gray} D.~F.,  2008b, \mn@doi [\aj] {10.1088/0004-6256/135/4/1450}, \href
  {https://ui.adsabs.harvard.edu/abs/2008AJ....135.1450G} {135, 1450}

\bibitem[\protect\citeauthoryear{{Gray} \& {Johanson}}{{Gray} \&
  {Johanson}}{1991}]{Gray1991}
{Gray} D.~F.,  {Johanson} H.~L.,  1991, \mn@doi [\pasp] {10.1086/132839}, \href
  {https://ui.adsabs.harvard.edu/abs/1991PASP..103..439G} {103, 439}

\bibitem[\protect\citeauthoryear{{Gustafsson}, {Edvardsson}, {Eriksson},
  {J{\o}rgensen}, {Nordlund}  \& {Plez}}{{Gustafsson}
  et~al.}{2008}]{Gustafsson2008}
{Gustafsson} B.,  {Edvardsson} B.,  {Eriksson} K.,  {J{\o}rgensen} U.~G.,
  {Nordlund} {\r{A}}.,   {Plez} B.,  2008, \mn@doi [\aap]
  {10.1051/0004-6361:200809724}, \href
  {https://ui.adsabs.harvard.edu/abs/2008A&A...486..951G} {486, 951}

\bibitem[\protect\citeauthoryear{{Hamano} et~al.,}{{Hamano}
  et~al.}{2015}]{Hamano2015}
{Hamano} S.,  et~al., 2015, \mn@doi [\apj] {10.1088/0004-637X/800/2/137}, \href
  {https://ui.adsabs.harvard.edu/abs/2015ApJ...800..137H} {800, 137}

\bibitem[\protect\citeauthoryear{{Haubois} et~al.,}{{Haubois}
  et~al.}{2009}]{Haubois2009}
{Haubois} X.,  et~al., 2009, \mn@doi [\aap] {10.1051/0004-6361/200912927},
  \href {https://ui.adsabs.harvard.edu/abs/2009A&A...508..923H} {508, 923}

\bibitem[\protect\citeauthoryear{{Heiter} \& {Eriksson}}{{Heiter} \&
  {Eriksson}}{2006}]{Heiter2006}
{Heiter} U.,  {Eriksson} K.,  2006, \mn@doi [\aap]
  {10.1051/0004-6361:20064925}, \href
  {https://ui.adsabs.harvard.edu/abs/2006A&A...452.1039H} {452, 1039}

\bibitem[\protect\citeauthoryear{{Heiter}, {Jofr{\'e}}, {Gustafsson}, {Korn},
  {Soubiran}  \& {Th{\'e}venin}}{{Heiter} et~al.}{2015}]{Heiter2015}
{Heiter} U.,  {Jofr{\'e}} P.,  {Gustafsson} B.,  {Korn} A.~J.,  {Soubiran} C.,
   {Th{\'e}venin} F.,  2015, \mn@doi [\aap] {10.1051/0004-6361/201526319},
  \href {http://adsabs.harvard.edu/abs/2015A%26A...582A..49H} {582, A49}

\bibitem[\protect\citeauthoryear{{Holtzman} et~al.,}{{Holtzman}
  et~al.}{2018}]{Holtzman2018}
{Holtzman} J.~A.,  et~al., 2018, \mn@doi [\aj] {10.3847/1538-3881/aad4f9},
  \href {https://ui.adsabs.harvard.edu/abs/2018AJ....156..125H} {156, 125}

\bibitem[\protect\citeauthoryear{{Huang}, {Wallerstein}  \& {Stone}}{{Huang}
  et~al.}{2012}]{Huang2012}
{Huang} W.,  {Wallerstein} G.,   {Stone} M.,  2012, \mn@doi [\aap]
  {10.1051/0004-6361/201219804}, \href
  {https://ui.adsabs.harvard.edu/abs/2012A&A...547A..62H} {547, A62}

\bibitem[\protect\citeauthoryear{{Ikeda} et~al.,}{{Ikeda}
  et~al.}{2016}]{Ikeda2016}
{Ikeda} Y.,  et~al., 2016, in Ground-based and Airborne Instrumentation for
  Astronomy VI. p. 99085Z, \mn@doi{10.1117/12.2230886}

\bibitem[\protect\citeauthoryear{{Ikeda} et~al.,}{{Ikeda}
  et~al.}{2018}]{Ikeda2018}
{Ikeda} Y.,  et~al., 2018, in \procspie. p. 107025U,
  \mn@doi{10.1117/12.2309605}

\bibitem[\protect\citeauthoryear{{Jian}, {Matsunaga}  \& {Fukue}}{{Jian}
  et~al.}{2019}]{Jian2019}
{Jian} M.,  {Matsunaga} N.,   {Fukue} K.,  2019, \mn@doi [\mnras]
  {10.1093/mnras/stz237}, \href
  {https://ui.adsabs.harvard.edu/abs/2019MNRAS.485.1310J} {485, 1310}

\bibitem[\protect\citeauthoryear{{Jian} et~al.,}{{Jian}
  et~al.}{2020}]{Jian2020}
{Jian} M.,  et~al., 2020, \mn@doi [\mnras] {10.1093/mnras/staa834}, \href
  {https://ui.adsabs.harvard.edu/abs/2020MNRAS.494.1724J} {494, 1724}

\bibitem[\protect\citeauthoryear{{Jofr{\'e}} et~al.,}{{Jofr{\'e}}
  et~al.}{2015}]{Jofre2015}
{Jofr{\'e}} P.,  et~al., 2015, \mn@doi [\aap] {10.1051/0004-6361/201526604},
  \href {https://ui.adsabs.harvard.edu/abs/2015A&A...582A..81J} {582, A81}

\bibitem[\protect\citeauthoryear{{Kervella}, {Verhoelst}, {Ridgway}, {Perrin},
  {Lacour}, {Cami}  \& {Haubois}}{{Kervella} et~al.}{2009}]{Kervella2009}
{Kervella} P.,  {Verhoelst} T.,  {Ridgway} S.~T.,  {Perrin} G.,  {Lacour} S.,
  {Cami} J.,   {Haubois} X.,  2009, \mn@doi [\aap]
  {10.1051/0004-6361/200912521}, \href
  {https://ui.adsabs.harvard.edu/abs/2009A&A...504..115K} {504, 115}

\bibitem[\protect\citeauthoryear{{Kiss}, {Szab{\'o}}  \& {Bedding}}{{Kiss}
  et~al.}{2006}]{Kiss2006}
{Kiss} L.~L.,  {Szab{\'o}} G.~M.,   {Bedding} T.~R.,  2006, \mn@doi [\mnras]
  {10.1111/j.1365-2966.2006.10973.x}, \href
  {https://ui.adsabs.harvard.edu/abs/2006MNRAS.372.1721K} {372, 1721}

\bibitem[\protect\citeauthoryear{{Koleva}, {Prugniel}, {Bouchard}  \&
  {Wu}}{{Koleva} et~al.}{2009}]{Koleva2009}
{Koleva} M.,  {Prugniel} P.,  {Bouchard} A.,   {Wu} Y.,  2009, \mn@doi [\aap]
  {10.1051/0004-6361/200811467}, \href
  {https://ui.adsabs.harvard.edu/abs/2009A&A...501.1269K} {501, 1269}

\bibitem[\protect\citeauthoryear{{Kondo} et~al.,}{{Kondo}
  et~al.}{2019}]{Kondo2019}
{Kondo} S.,  et~al., 2019, \mn@doi [\apj] {10.3847/1538-4357/ab0ec4}, \href
  {https://ui.adsabs.harvard.edu/abs/2019ApJ...875..129K} {875, 129}

\bibitem[\protect\citeauthoryear{{Kovalev}, {Bergemann}, {Ting}  \&
  {Rix}}{{Kovalev} et~al.}{2019}]{Kovalev2019}
{Kovalev} M.,  {Bergemann} M.,  {Ting} Y.-S.,   {Rix} H.-W.,  2019, \mn@doi
  [\aap] {10.1051/0004-6361/201935861}, \href
  {https://ui.adsabs.harvard.edu/abs/2019A&A...628A..54K} {628, A54}

\bibitem[\protect\citeauthoryear{{Kovtyukh}}{{Kovtyukh}}{2007}]{Kovtyukh2007}
{Kovtyukh} V.~V.,  2007, \mn@doi [\mnras] {10.1111/j.1365-2966.2007.11804.x},
  \href {https://ui.adsabs.harvard.edu/abs/2007MNRAS.378..617K} {378, 617}

\bibitem[\protect\citeauthoryear{{Kovtyukh}, {Soubiran}, {Bienaym{\'e}},
  {Mishenina}  \& {Belik}}{{Kovtyukh} et~al.}{2006}]{Kovtyukh2006}
{Kovtyukh} V.~V.,  {Soubiran} C.,  {Bienaym{\'e}} O.,  {Mishenina} T.~V.,
  {Belik} S.~I.,  2006, \mn@doi [\mnras] {10.1111/j.1365-2966.2006.10719.x},
  \href {https://ui.adsabs.harvard.edu/abs/2006MNRAS.371..879K} {371, 879}

\bibitem[\protect\citeauthoryear{{Kravchenko}, {Chiavassa}, {Van Eck},
  {Jorissen}, {Merle}, {Freytag}  \& {Plez}}{{Kravchenko}
  et~al.}{2019}]{Kravchenko2019}
{Kravchenko} K.,  {Chiavassa} A.,  {Van Eck} S.,  {Jorissen} A.,  {Merle} T.,
  {Freytag} B.,   {Plez} B.,  2019, \mn@doi [\aap]
  {10.1051/0004-6361/201935809}, \href
  {https://ui.adsabs.harvard.edu/abs/2019A&A...632A..28K} {632, A28}

\bibitem[\protect\citeauthoryear{{Kurucz}}{{Kurucz}}{2011}]{Kurucz2011}
{Kurucz} R.~L.,  2011, \mn@doi [Canadian Journal of Physics] {10.1139/p10-104},
  \href {https://ui.adsabs.harvard.edu/abs/2011CaJPh..89..417K} {89, 417}

\bibitem[\protect\citeauthoryear{{Ku{\v{c}}inskas} et~al.,}{{Ku{\v{c}}inskas}
  et~al.}{2013}]{Kucinskas2013}
{Ku{\v{c}}inskas} A.,  et~al., 2013, \mn@doi [\aap]
  {10.1051/0004-6361/201220240}, \href
  {https://ui.adsabs.harvard.edu/abs/2013A&A...549A..14K} {549, A14}

\bibitem[\protect\citeauthoryear{{Lambert}, {Brown}, {Hinkle}  \&
  {Johnson}}{{Lambert} et~al.}{1984}]{Lambert1984}
{Lambert} D.~L.,  {Brown} J.~A.,  {Hinkle} K.~H.,   {Johnson} H.~R.,  1984,
  \mn@doi [\apj] {10.1086/162401}, \href
  {https://ui.adsabs.harvard.edu/abs/1984ApJ...284..223L} {284, 223}

\bibitem[\protect\citeauthoryear{{Lan{\c{c}}on} \& {Hauschildt}}{{Lan{\c{c}}on}
  \& {Hauschildt}}{2010}]{Lancon2010}
{Lan{\c{c}}on} A.,  {Hauschildt} P.~H.,  2010, in {Leitherer} C.,  {Bennett}
  P.~D.,  {Morris} P.~W.,   {Van Loon} J.~T.,  eds,  Astronomical Society of
  the Pacific Conference Series Vol. 425, Hot and Cool: Bridging Gaps in
  Massive Star Evolution. p.~61

\bibitem[\protect\citeauthoryear{{Lan{\c{c}}on}, {Hauschildt}, {Ladjal}  \&
  {Mouhcine}}{{Lan{\c{c}}on} et~al.}{2007}]{Lancon2007}
{Lan{\c{c}}on} A.,  {Hauschildt} P.~H.,  {Ladjal} D.,   {Mouhcine} M.,  2007,
  \mn@doi [\aap] {10.1051/0004-6361:20065824}, \href
  {https://ui.adsabs.harvard.edu/abs/2007A&A...468..205L} {468, 205}

\bibitem[\protect\citeauthoryear{{Levesque} \& {Massey}}{{Levesque} \&
  {Massey}}{2020}]{Levesque2020}
{Levesque} E.~M.,  {Massey} P.,  2020, \mn@doi [\apjl]
  {10.3847/2041-8213/ab7935}, \href
  {https://ui.adsabs.harvard.edu/abs/2020ApJ...891L..37L} {891, L37}

\bibitem[\protect\citeauthoryear{{Levesque}, {Massey}, {Olsen}, {Plez},
  {Josselin}, {Maeder}  \& {Meynet}}{{Levesque} et~al.}{2005}]{Levesque2005}
{Levesque} E.~M.,  {Massey} P.,  {Olsen} K.~A.~G.,  {Plez} B.,  {Josselin} E.,
  {Maeder} A.,   {Meynet} G.,  2005, \mn@doi [\apj] {10.1086/430901}, \href
  {https://ui.adsabs.harvard.edu/abs/2005ApJ...628..973L} {628, 973}

\bibitem[\protect\citeauthoryear{{Levesque}, {Massey}, {Olsen}, {Plez},
  {Meynet}  \& {Maeder}}{{Levesque} et~al.}{2006}]{Levesque2006}
{Levesque} E.~M.,  {Massey} P.,  {Olsen} K.~A.~G.,  {Plez} B.,  {Meynet} G.,
  {Maeder} A.,  2006, \mn@doi [\apj] {10.1086/504417}, \href
  {https://ui.adsabs.harvard.edu/abs/2006ApJ...645.1102L} {645, 1102}

\bibitem[\protect\citeauthoryear{{Levesque}, {Massey}, {Olsen}  \&
  {Plez}}{{Levesque} et~al.}{2007}]{Levesque2007}
{Levesque} E.~M.,  {Massey} P.,  {Olsen} K.~A.~G.,   {Plez} B.,  2007, \mn@doi
  [\apj] {10.1086/520797}, \href
  {https://ui.adsabs.harvard.edu/abs/2007ApJ...667..202L} {667, 202}

\bibitem[\protect\citeauthoryear{{Lind}, {Bergemann}  \& {Asplund}}{{Lind}
  et~al.}{2012}]{Lind2012}
{Lind} K.,  {Bergemann} M.,   {Asplund} M.,  2012, \mn@doi [\mnras]
  {10.1111/j.1365-2966.2012.21686.x}, \href
  {https://ui.adsabs.harvard.edu/abs/2012MNRAS.427...50L} {427, 50}

\bibitem[\protect\citeauthoryear{{Lindegren} et~al.,}{{Lindegren}
  et~al.}{2021a}]{Lindegren2020a}
{Lindegren} L.,  et~al., 2021a, \mn@doi [\aap] {10.1051/0004-6361/202039709},
  \href {https://ui.adsabs.harvard.edu/abs/2021A&A...649A...2L} {649, A2}

\bibitem[\protect\citeauthoryear{{Lindegren} et~al.,}{{Lindegren}
  et~al.}{2021b}]{Lindegren2020b}
{Lindegren} L.,  et~al., 2021b, \mn@doi [\aap] {10.1051/0004-6361/202039653},
  \href {https://ui.adsabs.harvard.edu/abs/2021A&A...649A...4L} {649, A4}

\bibitem[\protect\citeauthoryear{{L{\'o}pez-Valdivia}
  et~al.,}{{L{\'o}pez-Valdivia} et~al.}{2019}]{LopezValdivia2019}
{L{\'o}pez-Valdivia} R.,  et~al., 2019, \mn@doi [\apj]
  {10.3847/1538-4357/ab2129}, \href
  {https://ui.adsabs.harvard.edu/abs/2019ApJ...879..105L} {879, 105}

\bibitem[\protect\citeauthoryear{{Luck} \& {Bond}}{{Luck} \&
  {Bond}}{1989}]{Luck1989}
{Luck} R.~E.,  {Bond} H.~E.,  1989, \mn@doi [\apjs] {10.1086/191386}, \href
  {https://ui.adsabs.harvard.edu/abs/1989ApJS...71..559L} {71, 559}

\bibitem[\protect\citeauthoryear{Markovsky \& Van~Huffel}{Markovsky \&
  Van~Huffel}{2007}]{Markovsky2007}
Markovsky I.,  Van~Huffel S.,  2007, Signal processing, 87, 2283

\bibitem[\protect\citeauthoryear{{Massey}}{{Massey}}{2003}]{Massey2003a}
{Massey} P.,  2003, \mn@doi [\araa] {10.1146/annurev.astro.41.071601.170033},
  \href {https://ui.adsabs.harvard.edu/abs/2003ARA&A..41...15M} {41, 15}

\bibitem[\protect\citeauthoryear{{Massey} \& {Evans}}{{Massey} \&
  {Evans}}{2016}]{Massey2016}
{Massey} P.,  {Evans} K.~A.,  2016, \mn@doi [\apj]
  {10.3847/0004-637X/826/2/224}, \href
  {https://ui.adsabs.harvard.edu/abs/2016ApJ...826..224M} {826, 224}

\bibitem[\protect\citeauthoryear{{Massey} \& {Olsen}}{{Massey} \&
  {Olsen}}{2003}]{Massey2003b}
{Massey} P.,  {Olsen} K.~A.~G.,  2003, \mn@doi [\aj] {10.1086/379558}, \href
  {https://ui.adsabs.harvard.edu/abs/2003AJ....126.2867M} {126, 2867}

\bibitem[\protect\citeauthoryear{{Massey}, {Plez}, {Levesque}, {Olsen},
  {Clayton}  \& {Josselin}}{{Massey} et~al.}{2005}]{Massey2005}
{Massey} P.,  {Plez} B.,  {Levesque} E.~M.,  {Olsen} K.~A.~G.,  {Clayton}
  G.~C.,   {Josselin} E.,  2005, \mn@doi [\apj] {10.1086/497065}, \href
  {https://ui.adsabs.harvard.edu/abs/2005ApJ...634.1286M} {634, 1286}

\bibitem[\protect\citeauthoryear{{Massey}, {Silva}, {Levesque}, {Plez},
  {Olsen}, {Clayton}, {Meynet}  \& {Maeder}}{{Massey}
  et~al.}{2009}]{Massey2009}
{Massey} P.,  {Silva} D.~R.,  {Levesque} E.~M.,  {Plez} B.,  {Olsen} K. A.~G.,
  {Clayton} G.~C.,  {Meynet} G.,   {Maeder} A.,  2009, \mn@doi [\apj]
  {10.1088/0004-637X/703/1/420}, \href
  {https://ui.adsabs.harvard.edu/abs/2009ApJ...703..420M} {703, 420}

\bibitem[\protect\citeauthoryear{{Matsunaga} et~al.,}{{Matsunaga}
  et~al.}{2020}]{Matsunaga2020}
{Matsunaga} N.,  et~al., 2020, \mn@doi [\apjs] {10.3847/1538-4365/ab5c25},
  \href {https://ui.adsabs.harvard.edu/abs/2020ApJS..246...10M} {246, 10}

\bibitem[\protect\citeauthoryear{{McWilliam}}{{McWilliam}}{1990}]{McWilliam1990}
{McWilliam} A.,  1990, \mn@doi [\apjs] {10.1086/191527}, \href
  {https://ui.adsabs.harvard.edu/abs/1990ApJS...74.1075M} {74, 1075}

\bibitem[\protect\citeauthoryear{{Montarg{\`e}s}, {Kervella}, {Perrin},
  {Ohnaka}, {Chiavassa}, {Ridgway}  \& {Lacour}}{{Montarg{\`e}s}
  et~al.}{2014}]{Montarges2014}
{Montarg{\`e}s} M.,  {Kervella} P.,  {Perrin} G.,  {Ohnaka} K.,  {Chiavassa}
  A.,  {Ridgway} S.~T.,   {Lacour} S.,  2014, \mn@doi [\aap]
  {10.1051/0004-6361/201423538}, \href
  {https://ui.adsabs.harvard.edu/abs/2014A&A...572A..17M} {572, A17}

\bibitem[\protect\citeauthoryear{{Neugent}, {Massey}, {Skiff}  \&
  {Meynet}}{{Neugent} et~al.}{2012}]{Neugent2012}
{Neugent} K.~F.,  {Massey} P.,  {Skiff} B.,   {Meynet} G.,  2012, \mn@doi
  [\apj] {10.1088/0004-637X/749/2/177}, \href
  {https://ui.adsabs.harvard.edu/abs/2012ApJ...749..177N} {749, 177}

\bibitem[\protect\citeauthoryear{{Obrien} \& {Lambert}}{{Obrien} \&
  {Lambert}}{1986}]{Obrien1986}
{Obrien} George~T. J.,  {Lambert} D.~L.,  1986, \mn@doi [\apjs]
  {10.1086/191160}, \href
  {https://ui.adsabs.harvard.edu/abs/1986ApJS...62..899O} {62, 899}

\bibitem[\protect\citeauthoryear{{Oestreicher} \&
  {Schmidt-Kaler}}{{Oestreicher} \& {Schmidt-Kaler}}{1998}]{Oestreicher1998}
{Oestreicher} M.~O.,  {Schmidt-Kaler} T.,  1998, \mn@doi [\mnras]
  {10.1046/j.1365-8711.1998.01501.x}, \href
  {https://ui.adsabs.harvard.edu/abs/1998MNRAS.299..625O} {299, 625}

\bibitem[\protect\citeauthoryear{{Ohnaka} et~al.,}{{Ohnaka}
  et~al.}{2011}]{Ohnaka2011}
{Ohnaka} K.,  et~al., 2011, \mn@doi [\aap] {10.1051/0004-6361/201016279}, \href
  {https://ui.adsabs.harvard.edu/abs/2011A&A...529A.163O} {529, A163}

\bibitem[\protect\citeauthoryear{{Ohnaka}, {Hofmann}, {Schertl}, {Weigelt},
  {Baffa}, {Chelli}, {Petrov}  \& {Robbe-Dubois}}{{Ohnaka}
  et~al.}{2013}]{Ohnaka2013}
{Ohnaka} K.,  {Hofmann} K.~H.,  {Schertl} D.,  {Weigelt} G.,  {Baffa} C.,
  {Chelli} A.,  {Petrov} R.,   {Robbe-Dubois} S.,  2013, \mn@doi [\aap]
  {10.1051/0004-6361/201321063}, \href
  {https://ui.adsabs.harvard.edu/abs/2013A&A...555A..24O} {555, A24}

\bibitem[\protect\citeauthoryear{{Origlia} et~al.,}{{Origlia}
  et~al.}{2019}]{Origlia2019}
{Origlia} L.,  et~al., 2019, \mn@doi [\aap] {10.1051/0004-6361/201936283},
  \href {https://ui.adsabs.harvard.edu/abs/2019A&A...629A.117O} {629, A117}

\bibitem[\protect\citeauthoryear{{Pasquato}, {Pourbaix}  \&
  {Jorissen}}{{Pasquato} et~al.}{2011}]{Pasquato2011}
{Pasquato} E.,  {Pourbaix} D.,   {Jorissen} A.,  2011, \mn@doi [\aap]
  {10.1051/0004-6361/201116859}, \href
  {https://ui.adsabs.harvard.edu/abs/2011A&A...532A..13P} {532, A13}

\bibitem[\protect\citeauthoryear{{Patrick}, {Evans}, {Davies}, {Kudritzki},
  {Ferguson}, {Bergemann}, {Pietrzy{\'n}ski}  \& {Turner}}{{Patrick}
  et~al.}{2017}]{Patrick2017}
{Patrick} L.~R.,  {Evans} C.~J.,  {Davies} B.,  {Kudritzki} R.~P.,  {Ferguson}
  A.~M.~N.,  {Bergemann} M.,  {Pietrzy{\'n}ski} G.,   {Turner} O.,  2017,
  \mn@doi [\mnras] {10.1093/mnras/stx410}, \href
  {https://ui.adsabs.harvard.edu/abs/2017MNRAS.468..492P} {468, 492}

\bibitem[\protect\citeauthoryear{{Prugniel}, {Vauglin}  \& {Koleva}}{{Prugniel}
  et~al.}{2011}]{Prugniel2011}
{Prugniel} P.,  {Vauglin} I.,   {Koleva} M.,  2011, \mn@doi [\aap]
  {10.1051/0004-6361/201116769}, \href
  {http://adsabs.harvard.edu/abs/2011A%26A...531A.165P} {531, A165}

\bibitem[\protect\citeauthoryear{{Ren}, {Jiang}, {Yang}  \& {Gao}}{{Ren}
  et~al.}{2019}]{Ren2019}
{Ren} Y.,  {Jiang} B.-W.,  {Yang} M.,   {Gao} J.,  2019, \mn@doi [\apjs]
  {10.3847/1538-4365/ab0825}, \href
  {https://ui.adsabs.harvard.edu/abs/2019ApJS..241...35R} {241, 35}

\bibitem[\protect\citeauthoryear{{Ryabchikova}, {Piskunov}, {Kurucz},
  {Stempels}, {Heiter}, {Pakhomov}  \& {Barklem}}{{Ryabchikova}
  et~al.}{2015}]{Ryabchikova2015}
{Ryabchikova} T.,  {Piskunov} N.,  {Kurucz} R.~L.,  {Stempels} H.~C.,  {Heiter}
  U.,  {Pakhomov} Y.,   {Barklem} P.~S.,  2015, \mn@doi [\physscr]
  {10.1088/0031-8949/90/5/054005}, \href
  {https://ui.adsabs.harvard.edu/abs/2015PhyS...90e4005R} {90, 054005}

\bibitem[\protect\citeauthoryear{{Sameshima} et~al.,}{{Sameshima}
  et~al.}{2018}]{Sameshima2018}
{Sameshima} H.,  et~al., 2018, \mn@doi [\pasp] {10.1088/1538-3873/aac1b4},
  \href {https://ui.adsabs.harvard.edu/abs/2018PASP..130g4502S} {130, 074502}

\bibitem[\protect\citeauthoryear{{S{\'a}nchez-Bl{\'a}zquez}
  et~al.,}{{S{\'a}nchez-Bl{\'a}zquez} et~al.}{2006}]{SanchezBlazquez2006}
{S{\'a}nchez-Bl{\'a}zquez} P.,  et~al., 2006, \mn@doi [\mnras]
  {10.1111/j.1365-2966.2006.10699.x}, \href
  {https://ui.adsabs.harvard.edu/abs/2006MNRAS.371..703S} {371, 703}

\bibitem[\protect\citeauthoryear{{Scargle} \& {Strecker}}{{Scargle} \&
  {Strecker}}{1979}]{Scargle1979}
{Scargle} J.~D.,  {Strecker} D.~W.,  1979, \mn@doi [\apj] {10.1086/156910},
  \href {https://ui.adsabs.harvard.edu/abs/1979ApJ...228..838S} {228, 838}

\bibitem[\protect\citeauthoryear{{Schultz} \& {Armentrout}}{{Schultz} \&
  {Armentrout}}{1991}]{Schultz1991}
{Schultz} R.~H.,  {Armentrout} P.~B.,  1991, \mn@doi [\jcp] {10.1063/1.459897},
  \href {https://ui.adsabs.harvard.edu/abs/1991JChPh..94.2262S} {94, 2262}

\bibitem[\protect\citeauthoryear{{Skrutskie} et~al.,}{{Skrutskie}
  et~al.}{2006}]{Skrutskie2006}
{Skrutskie} M.~F.,  et~al., 2006, \mn@doi [\aj] {10.1086/498708}, \href
  {https://ui.adsabs.harvard.edu/abs/2006AJ....131.1163S} {131, 1163}

\bibitem[\protect\citeauthoryear{{Smartt}}{{Smartt}}{2015}]{Smartt2015}
{Smartt} S.~J.,  2015, \mn@doi [\pasa] {10.1017/pasa.2015.17}, \href
  {https://ui.adsabs.harvard.edu/abs/2015PASA...32...16S} {32, e016}

\bibitem[\protect\citeauthoryear{{Sneden}}{{Sneden}}{1973}]{Sneden1973}
{Sneden} C.,  1973, \mn@doi [\apj] {10.1086/152374}, \href
  {https://ui.adsabs.harvard.edu/abs/1973ApJ...184..839S} {184, 839}

\bibitem[\protect\citeauthoryear{{Sneden}, {Bean}, {Ivans}, {Lucatello}  \&
  {Sobeck}}{{Sneden} et~al.}{2012}]{Sneden2012}
{Sneden} C.,  {Bean} J.,  {Ivans} I.,  {Lucatello} S.,   {Sobeck} J.,  2012,
  {MOOG: LTE line analysis and spectrum synthesis} (\mn@eprint {ascl}
  {1202.009})

\bibitem[\protect\citeauthoryear{{Soraisam} et~al.,}{{Soraisam}
  et~al.}{2018}]{Soraisam2018}
{Soraisam} M.~D.,  et~al., 2018, \mn@doi [\apj] {10.3847/1538-4357/aabc59},
  \href {https://ui.adsabs.harvard.edu/abs/2018ApJ...859...73S} {859, 73}

\bibitem[\protect\citeauthoryear{{Strassmeier} \& {Schordan}}{{Strassmeier} \&
  {Schordan}}{2000}]{Strassmeier2000}
{Strassmeier} K.~G.,  {Schordan} P.,  2000, \mn@doi [Astronomische Nachrichten]
  {10.1002/1521-3994(200012)321:5/6<277::AID-ASNA277>3.0.CO;2-H}, \href
  {https://ui.adsabs.harvard.edu/abs/2000AN....321..277S} {321, 277}

\bibitem[\protect\citeauthoryear{{Tabernero}, {Dorda}, {Negueruela}  \&
  {Gonz{\'a}lez-Fern{\'a}ndez}}{{Tabernero} et~al.}{2018}]{Tabernero2018}
{Tabernero} H.~M.,  {Dorda} R.,  {Negueruela} I.,
  {Gonz{\'a}lez-Fern{\'a}ndez} C.,  2018, \mn@doi [\mnras]
  {10.1093/mnras/sty399}, \href
  {https://ui.adsabs.harvard.edu/abs/2018MNRAS.476.3106T} {476, 3106}

\bibitem[\protect\citeauthoryear{{Taniguchi} et~al.,}{{Taniguchi}
  et~al.}{2018}]{Taniguchi2018}
{Taniguchi} D.,  et~al., 2018, \mn@doi [\mnras] {10.1093/mnras/stx2691}, \href
  {https://ui.adsabs.harvard.edu/abs/2018MNRAS.473.4993T} {473, 4993}

\bibitem[\protect\citeauthoryear{{Teixeira}, {Sousa}, {Tsantaki}, {Monteiro},
  {Santos}  \& {Israelian}}{{Teixeira} et~al.}{2016}]{Teixeira2016}
{Teixeira} G.~D.~C.,  {Sousa} S.~G.,  {Tsantaki} M.,  {Monteiro}
  M.~J.~P.~F.~G.,  {Santos} N.~C.,   {Israelian} G.,  2016, \mn@doi [\aap]
  {10.1051/0004-6361/201525783}, \href
  {https://ui.adsabs.harvard.edu/abs/2016A&A...595A..15T} {595, A15}

\bibitem[\protect\citeauthoryear{{Torra} et~al.,}{{Torra}
  et~al.}{2021}]{Torra2020}
{Torra} F.,  et~al., 2021, \mn@doi [\aap] {10.1051/0004-6361/202039637}, \href
  {https://ui.adsabs.harvard.edu/abs/2021A&A...649A..10T} {649, A10}

\bibitem[\protect\citeauthoryear{{Tsuji}}{{Tsuji}}{1976}]{Tsuji1976}
{Tsuji} T.,  1976, \pasj, \href
  {https://ui.adsabs.harvard.edu/abs/1976PASJ...28..567T} {28, 567}

\bibitem[\protect\citeauthoryear{{Tsuji}}{{Tsuji}}{2000}]{Tsuji2000}
{Tsuji} T.,  2000, \mn@doi [\apj] {10.1086/309185}, \href
  {https://ui.adsabs.harvard.edu/abs/2000ApJ...538..801T} {538, 801}

\bibitem[\protect\citeauthoryear{{Tsuji}}{{Tsuji}}{2006}]{Tsuji2006}
{Tsuji} T.,  2006, \mn@doi [\apj] {10.1086/504585}, \href
  {https://ui.adsabs.harvard.edu/abs/2006ApJ...645.1448T} {645, 1448}

\bibitem[\protect\citeauthoryear{{Walmswell} \& {Eldridge}}{{Walmswell} \&
  {Eldridge}}{2012}]{Walmswell2012}
{Walmswell} J.~J.,  {Eldridge} J.~J.,  2012, \mn@doi [\mnras]
  {10.1111/j.1365-2966.2011.19860.x}, \href
  {https://ui.adsabs.harvard.edu/abs/2012MNRAS.419.2054W} {419, 2054}

\bibitem[\protect\citeauthoryear{{Wasatonic}, {Guinan}  \&
  {Durbin}}{{Wasatonic} et~al.}{2015}]{Wasatonic2015}
{Wasatonic} R.~P.,  {Guinan} E.~F.,   {Durbin} A.~J.,  2015, \mn@doi [\pasp]
  {10.1086/683261}, \href
  {https://ui.adsabs.harvard.edu/abs/2015PASP..127.1010W} {127, 1010}

\bibitem[\protect\citeauthoryear{{Wenger} et~al.,}{{Wenger}
  et~al.}{2000}]{Wenger2000}
{Wenger} M.,  et~al., 2000, \mn@doi [\aaps] {10.1051/aas:2000332}, \href
  {http://adsabs.harvard.edu/abs/2000A%26AS..143....9W} {143, 9}

\bibitem[\protect\citeauthoryear{{White} \& {Wing}}{{White} \&
  {Wing}}{1978}]{White1978}
{White} N.~M.,  {Wing} R.~F.,  1978, \mn@doi [\apj] {10.1086/156136}, \href
  {https://ui.adsabs.harvard.edu/abs/1978ApJ...222..209W} {222, 209}

\bibitem[\protect\citeauthoryear{{Wittkowski}, {Arroyo-Torres}, {Marcaide},
  {Abellan}, {Chiavassa}  \& {Guirado}}{{Wittkowski}
  et~al.}{2017}]{Wittkowski2017}
{Wittkowski} M.,  {Arroyo-Torres} B.,  {Marcaide} J.~M.,  {Abellan} F.~J.,
  {Chiavassa} A.,   {Guirado} J.~C.,  2017, \mn@doi [\aap]
  {10.1051/0004-6361/201629349}, \href
  {https://ui.adsabs.harvard.edu/abs/2017A&A...597A...9W} {597, A9}

\bibitem[\protect\citeauthoryear{{Wu}, {Singh}, {Prugniel}, {Gupta}  \&
  {Koleva}}{{Wu} et~al.}{2011}]{Wu2011}
{Wu} Y.,  {Singh} H.~P.,  {Prugniel} P.,  {Gupta} R.,   {Koleva} M.,  2011,
  \mn@doi [\aap] {10.1051/0004-6361/201015014}, \href
  {https://ui.adsabs.harvard.edu/abs/2011A&A...525A..71W} {525, A71}

\bibitem[\protect\citeauthoryear{{Yang} et~al.,}{{Yang}
  et~al.}{2018}]{Yang2018}
{Yang} M.,  et~al., 2018, \mn@doi [\aap] {10.1051/0004-6361/201832833}, \href
  {https://ui.adsabs.harvard.edu/abs/2018A&A...616A.175Y} {616, A175}

\bibitem[\protect\citeauthoryear{{van Leeuwen}}{{van
  Leeuwen}}{2007}]{vanLeeuwen2007}
{van Leeuwen} F.,  2007, \mn@doi [\aap] {10.1051/0004-6361:20078357}, \href
  {https://ui.adsabs.harvard.edu/abs/2007A&A...474..653V} {474, 653}

\makeatother
\end{thebibliography}

\bsp 
\label{lastpage}
\end{document}